\documentclass[10pt]{article}
\usepackage{amsmath}
\usepackage{amsfonts}
\usepackage{amssymb}
\usepackage{mathrsfs}
\usepackage{fancyhdr}
\usepackage{blindtext}
\usepackage{MnSymbol}
\usepackage[cal=boondox,scr=boondoxo]{mathalfa}
\usepackage{float}
\usepackage{subfig}
\usepackage{fancybox,graphicx}
\usepackage{subfig}
\usepackage{caption}
\usepackage{color}
\usepackage{authblk}
\usepackage[colorlinks]{hyperref}
\usepackage{accents}
\usepackage[titletoc,title]{appendix}
\usepackage{cite}
\usepackage{mathtools}
\usepackage[top=1in, bottom=1in, left=1in, right=1in]{geometry}
\usepackage{stackengine}
\usepackage{bm}
\usepackage{lineno}
\usepackage{footnote}
\usepackage[utf8]{inputenc}

\begin{document}
\pdfoutput=1
\captionsetup{font={footnotesize}}

\title{Quantum-inspired Beamforming Optimization for Quantized Phase-only Massive MIMO Arrays} 
 
\author[1,\ $\ast$]{Yutong Jiang}
\author[1,\ $\ast$]{Hangyu Ge}
\author[3,\ $\dagger$]{Bi-Ying Wang}
\author[1]{Shuai S. A. Yuan}
\author[4]{Shi-Jie Pan}
\author[5]{Hongjing Xu}
\author[6]{Xiaopeng Cui}
\author[2]{Man-Hong Yung}
\author[1]{Feng Liu}
\author[1,\ $\dagger$]{Wei E. I. Sha}

\affil[1]{Zhejiang University, Hangzhou 310027, China.}
\affil[2]{Shenzhen Institute for Quantum Science and Engineering, Southern University of Science and Technology, Shenzhen, China.}
\affil[3]{Yangtze River Delta Industrial Innovation Center of Quantum Science and Technology, Suzhou, China.}
\affil[4]{State Key Laboratory of Networking and Switching Technology, Beijing University of Posts and Telecommunications, Beijing 100876, China.}
\affil[5]{Department of Electronic Engineering Beijing National Research Center for Information Science and Technology, Tsinghua University, Beijing, China.}
\affil[6]{Department of Physics and State Key Laboratory of Surface Physics, Fudan University, Shanghai, China.}

\maketitle

\renewcommand{\thefootnote}{\fnsymbol{footnote}} 
\footnotetext[1]{\ These authors contributed equally to this work.}
\footnotetext[2]{\ Email to: biying@mail.ustc.edu.cn,\ weisha@zju.edu.cn.}

\begin{abstract}
This paper introduces an innovative quantum-inspired method for beamforming (BF) optimization in multiple-input multiple-output (MIMO) arrays. The method leverages the simulated bifurcation (SB) algorithm to address the complex combinatorial optimization problem due to the quantized phase configuration. We propose novel encoding techniques for phase quantization, which are then mapped into Ising spins. This enables efficient construction of the Hamiltonian and subsequent optimization of BF patterns. The results clearly demonstrate that the SB optimizer surpasses traditional schemes such as digital BF, holographic algorithm, simulated annealing algorithm and genetic algorithm, offering shorter optimization time and higher solution quality with reliability. The impressive capability of the SB optimizer to handle complex BF scenarios, including wide-angle sidelobe suppression and multiple beams with nulls, is undoubtedly demonstrated through several application cases. These findings strongly suggest that quantum-inspired methods have great potential to advance MIMO techniques in next-generation wireless communication. 
\end{abstract}

\section{Introduction}
Multiple-input and multiple-output (MIMO) technology, that is, the use of multiple antennas at the transmitter and receiver, has attracted much attention in wireless communications. Advances in the MIMO technology result in high data rates, strong signal reliability and large network capacity\cite{lu2014overview,paulraj2004overview,hampton2013introduction,tsoulos2018mimo,goldsmith2003capacity,haimovich2007mimo}. 
Moreover, MIMO technology encompasses a variety of powerful techniques, such as beamforming (BF), that significantly enhance the performance of wireless communication\cite{ji2024electromagnetic,kim2013multi,sun2014mimo,kulkarni2016comparison}.
While MIMO technology has been explored for more than a decade, the seminal work of Marzetta introduced the exciting new concept of ``massive MIMO", where the number of antenna elements at the base station reaches dozens or hundreds\cite{marzetta2015massive}. 
Massive MIMO arrays allow for more concentrated beams directing toward the users, which enables enhanced signal quality, extended coverage and increased capacity\cite{larsson2014massive,marzetta2016fundamentals,bjornson2016massive}.
Massive MIMO BF is particularly useful to obtain desired array gains, thereby offering both increased signal-to-noise ratio and additional radio link margin that mitigates propagation path loss. 
This promising technology is expected to play a critical role in 5G mobile systems\cite{molisch2017hybrid,ali2017beamforming,wu2018hybrid,yang2018digital,maksymyuk2018deep}. Furthermore, to reduce hardware costs, most MIMO arrays adopt a quantized phase-only configuration, where the excitation amplitude of each element remains the same, eliminating the need for amplifiers and attenuators. In phase-only optimization, hardware complexity and deployment difficulty are significantly reduced and there is no need to consider power-related constraints. Quantized phase-only massive MIMO BF needs to address emerging complex scenarios that demand high precision and flexibility. For example, in low-altitude economy applications, such as drone swarms and monitoring networks, complex tasks like deep nulls are required to mitigate interference. Another example is satellite-ground coordination, where BF with wide-angle sidelobe suppression is essential to avoid interference with satellite links.

However, a large number of quantized phase-only elements in massive MIMO arrays poses major challenges. Specifically, quantized phase-only BF can be mapped into an unconstrained combinatorial optimization problem, which is NP-hard. To address these challenges, many BF optimization schemes have been introduced in the literature, which can be broadly categorized into four classes:
\textbf{(Semi-)analytical schemes:} Approaches such as holographic BF leverage holographic principles to optimize beam patterns\cite{black2017holographic}. While computationally efficient for simple scenarios such as single BF or multiple beams, they struggle with complex objectives like multiple beams with nulls and are prone to phase quantization errors.
\textbf{Constrained optimization-based schemes:} Techniques like digital BF (DBF) with Constrained Minimum Variance (CMV) utilize convex optimization to control phase and amplitude at each antenna element\cite{yang2018digital, litva1996digital}. Although they offer high performance for specific applications, they are limited by phase quantization errors and performance degradation when discarding amplitude information to retain only phase.
\textbf{Stochastic optimization schemes:} Methods such as genetic algorithms (GA) and simulated annealing (SA) are inspired by natural processes and are capable of handling non-linear optimization problems\cite{guo2017genetic,SA_2019_Ismayilov}. However, they suffer from slow convergence and a tendency to fall into local optima.
\textbf{Machine learning schemes:} Approaches such as neural networks show promise but require extensive training data and not generalize well to new scenarios or complex objectives\cite{ahmed2021machine,liu2020machine}.

Despite these efforts, the optimization of quantized phase-only massive MIMO for BF in complex scenarios remains a critical challenge, primarily due to two key issues: \textbf{prolonged optimization time} and \textbf{compromised solution quality}.
The massive MIMO architecture, particularly with high precision quantization, leads to an exponentially large search space, which significantly increases the optimization time and hinders its applicability in real-world scenarios\cite{haupt1997phase,deford1988phase,morabito2012effective}.
Moreover, the constraints imposed by quantization and phase-only operations often force non-optimization algorithms, such as the holographic BF and digital BF, to discard amplitude information or approximate neighboring phases, resulting in suboptimal BF solutions.
Additionally, in massive MIMO systems, the simultaneous optimization for large-scale arrays in complex scenarios —characterized by intricate objective functions (e.g., directivity and sidelobe levels)—dramatically increases the complexity of the energy landscape. This complexity often traps the optimization process in local optima, preventing the attainment of globally optimal solutions.
Particularly, these challenges are further compounded by demands for scalability and computational efficiency in large-scale MIMO systems.

Recently, quantum annealing has been proposed to solve EM problems and demonstrated its potential to obtain high-quality solutions in a short optimization time\cite{zeng2024performance,krikidis2024optimizing,de2021materials,motta2022emerging,hidary2019quantum}.
The Ising model for reconfigurable intelligent surfaces was first introduced and analyzed in 2021\cite{ross2021engineering}, in which the BF Hamiltonian represents the total scattered power of the reconfigurable intelligent surface along a specific direction. The 1-bit and 2-bit quantized phase-only BF cases for small-scale reconfigurable intelligent surfaces are shown, confirming the effectiveness of quantum annealing.
However, though promising for combinatorial optimization, quantum annealing platforms encounter inherent scalability limitations when addressing fully connected problems like MIMO BF. The quantum annealer suffers from restricted qubit connectivity, requiring multiple physical qubits to emulate a fully interconnected logical variable\cite{boothby2016fast}. For instance, while the D-Wave Advantage platform (Pegasus architecture) integrates 5,000+ qubits, their practical capacity for fully connected variables remains limited to hundreds ($<$ 200)\cite{boothby2020next,willsch2022benchmarking}. Therefore, current hardware struggles to directly deal with the large-scale MIMO BF optimization.

Inspired by the fundamental principle of quantum annealing, in this work, we utilize a quantum-inspired method called simulated bifurcation (SB) algorithm\cite{goto2016bifurcation,goto2019combinatorial,goto2021high, Biyingwang2023quantum,obata2024ultra}, which simulates the adiabatic evolution of a classical nonlinear oscillator on a classical computer. The SB algorithm provides an efficient approach to solving the BF optimization problem in quantized phase-only massive MIMO arrays by mapping a nonlinear quantum model onto a classical framework and solving the resulting equations of motion on a classical computer. This unique combination of quantum principles and classical computation leverages quantum-inspired characteristics to significantly enhance both the speed and quality of the solution, while also addressing the scalability challenges to effectively navigate the complex energy landscapes of massive MIMO systems, ensuring robust and scalable solutions for real-world applications.
The development of our approach involves three key steps. \textbf{Innovative encoding technique}: We propose a novel encoding method to map the phases into sets of spins, enabling efficient optimization of 3-bit phase distributions. \textbf{Hamiltonian formulatio}n: We formulate the BF optimization Hamiltonian as a variable-weighted sum of the target Hamiltonian and initial nonlinear Kerr Hamiltonian. This formulation captures the complex objectives of BF optimization while maintaining computational tractability. \textbf{Quantum-inspired optimization}: We implement the SB algorithm on a classical computer to simulate the adiabatic evolution of the Hamiltonian, achieving high-quality solutions with significantly reduced optimization time.

In this work, we address the critical challenges of quantized phase-only massive MIMO BF optimization in complex scenarios, focusing on optimization time and solution quality. Our contributions are threefold:
\begin{itemize}
    \item \textbf{Problem Formulation:} We identify and tackle the key issues of large search spaces due to high precision phase quantization, phase-only constraints that degrade solution quality, and complex energy landscapes arising from complicated objective functions, which hinder the efficiency and performance of BF optimization in massive MIMO system wireless communication.
    \item \textbf{Technical Innovation:} We propose a quantum-inspired solution using the Simulated Bifurcation (SB) algorithm, combining a novel phase mapping encoding technique and a weighted Hamiltonian formulation to efficiently solve the BF optimization problem \textbf{for the first time}.
    \item \textbf{Numerical Experimental Validation:} We validate our approach through extensive experiments in three real-world scenarios, demonstrating its superiority over holographic algorithm, digital BF with constrained minimum variance, and optimization methods (SA, GA) in speed or quality. Additionally, we explore the parameters and characteristics of the SB algorithm, providing insights into its robustness and scalability, and the precision-performance tradeoff across different encoding schemes. 
\end{itemize}

\section{Principles}
\subsection{Background}
\subsubsection{Problem Statement} 
\begin{figure*}[htbp]
\centering 
\includegraphics[width=0.9\textwidth]{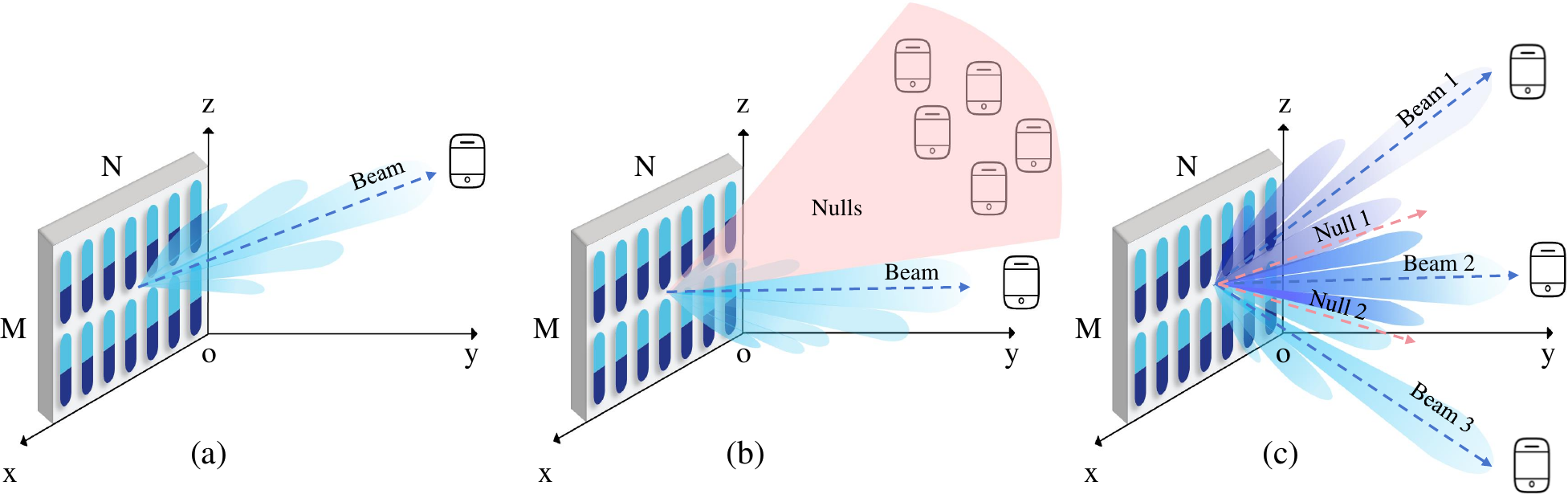}
    \caption[The system.]{Three common communication scenarios using a MIMO antenna array. (a) Single BF towards the user. (b) Single BF with sidelobe suppression. The red sector covering non-target users represents a wide range of pre-defined nulls. (c) Multiple BF with pre-defined nulls. Generating multiple beams towards several users with nulls to ensure the communication quality.}
\label{Fig1_problem_statement}
\vspace{-0.5\baselineskip}
\end{figure*}

The communication scenarios between a base station and users are shown in Fig. \ref{Fig1_problem_statement}. The base station is equipped with an antenna array composed of a large number of antenna elements.
Fig. \ref{Fig1_problem_statement}(a) shows a simple case that the array achieves BF in a single direction towards the user. For a more complex case shown in Fig. \ref{Fig1_problem_statement}(b), there are many other users around the target user distributed in a sector. To avoid beam interference in the wide range of non-target areas, sidelobe suppression is required. For Fig. \ref{Fig1_problem_statement}(c), space-division multiplexing is realized by forming multiple beams. Apart from BF, nulls are achieved in pre-defined directions to avoid interference among different users.

Each element in the antenna array has several quantized phases.These discrete phases are typically uniformly distributed across the full $2\pi$ range, with the quantization resolution directly determined by the encoding precision. For instance, when employing a 1-bit encoding scheme, the phase of each element will be either precisely 0 or ${\pi}$. 
\subsubsection{Binary Spin Model} 
\begin{figure}[htbp]
\centering 
\includegraphics[width=0.45\textwidth]{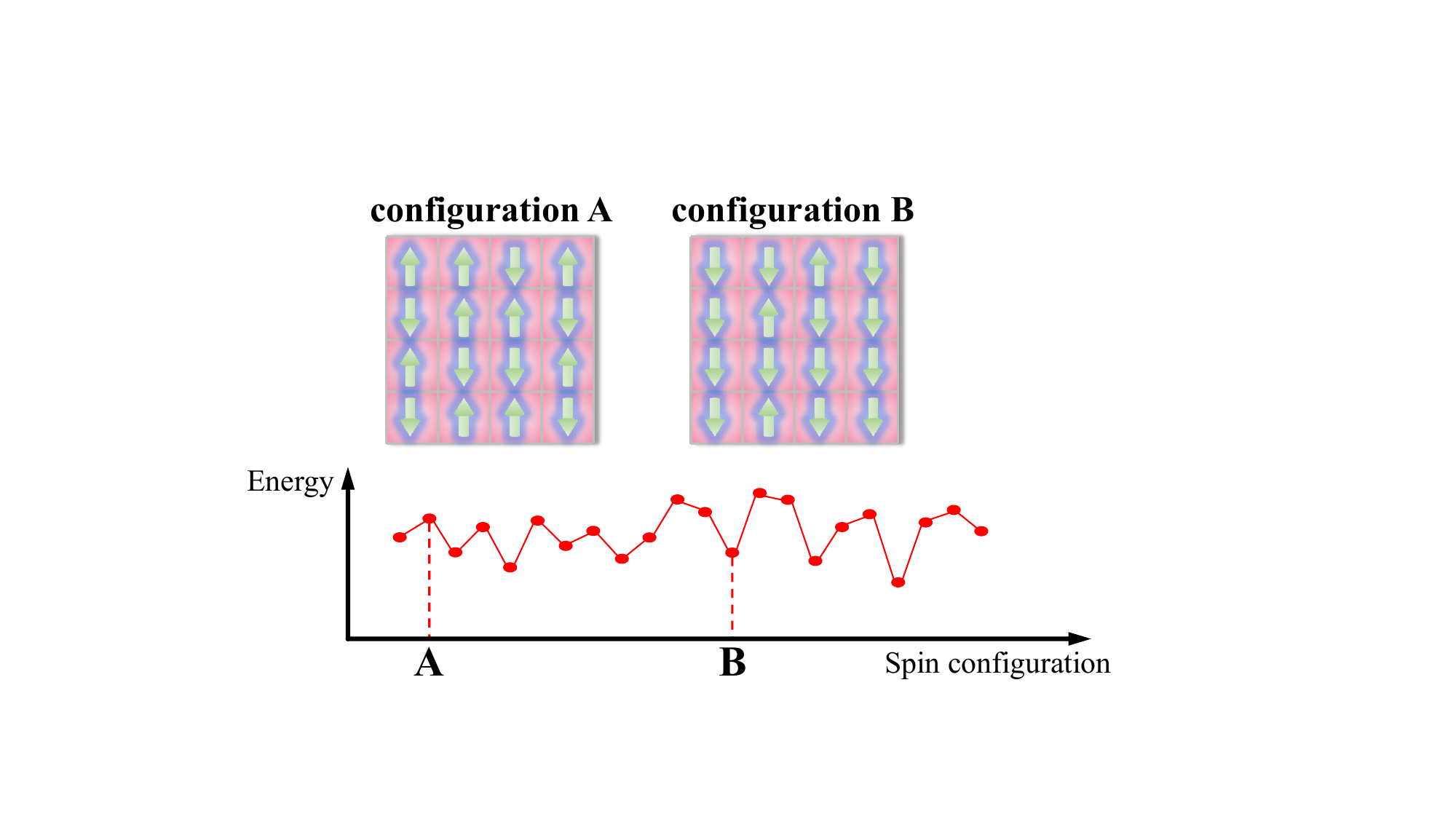}
    \caption[The system.]{Schematic diagram of the Ising model. Ising spins are located in lattices, and the energy of the system depends on the spin configuration.}
\label{Ising model}
\end{figure}

Due to phase discretization, BF can be formulated as a combinatorial optimization problem. For 1-bit and 2-bit encoding, the highest order of spin-spin interaction is 2, and BF is converted to a quadratic unconstrained binary optimization (QUBO) problem \cite{zaman2021pyqubo,glover2018tutorial,pastorello2019quantum}. The Ising model, a foundational framework for solving QUBO problems, consists of spins with two possible states: spin up (+1) and spin down (-1). These spins interact within a periodic structure, and the energy of the system is determined by their configuration \cite{schultz1964two}, as shown in Fig.~\ref{Ising model}. The Hamiltonian of a QUBO problem based on a 2D Ising spin model with $N_{s}$ spins is given by:

\begin{equation}
H_{QUBO} = {\sum\limits_{i = 1}^{N_{s}}{\sum\limits_{j = 1}^{N_{s}}{J_{ij}s_{i}s_{j}}}},
\label{Ising Hamiltonian}
\end{equation}
where $s_i$ and $s_j$ represent the states of the $i^{th}$ and $j^{th}$ spins, respectively, which can take values of either +1 (spin up) or -1 (spin down). $J_{ij}$ denotes the interaction strength between the spins $i$ and $j$, which determines the energy contribution of their interaction to the Hamiltonian.

While the Ising model is effective for problems with up to second-order interactions, it cannot handle higher-order interactions. For such cases, we employ higher-order unconstrained binary optimization (HUBO), which includes interactions among three or more spins. For instance, the Hamiltonian of a HUBO problem could be expressed as:

\begin{equation}
H_{HUBO} = {\sum\limits_{i = 1}^{N_{s}}{\sum\limits_{j = 1}^{N_{s}}{J_{ij}s_{i}s_{j}}}} + {\sum\limits_{i = 1}^{N_{s}}{\sum\limits_{j = 1}^{N_{s}}{\sum\limits_{p = 1}^{N_{s}}{J_{ijp}s_{i}s_{j}s_{p}}}}},
\label{HUBO Hamiltonian}
\end{equation}
where $J_{ijp} = 0$ when at least two of $i$, $j$, and $p$ are equal.  

The process of solving a combinatorial optimization problem involves the following steps: firstly, mapping the optimization variables to spins; secondly, defining the objective function; thirdly, substituting the spins into the objective function to express it as a Hamiltonian; last, finding the spin configuration that minimizes the Hamiltonian.

\subsection{Encoding Technique}
Unlike conventional approaches that represent discretized phases through linear superposition of spins, we introduce high-order terms into the phase expression, achieving higher-precision phase quantization. This encoding technique expands the solution space, generalizes to arbitrary-bit scenarios, and holds significant value in complex BF problems. In the \textit{$N_{b}$}-bit phase encoding, the phase of each element is quantized into $2^{N_{b}}$ discrete values within the range of 0 to 2$\pi$ and the phase expression of element (\textit{m},\textit{n}) is:
\begin{equation}
\begin{aligned}
e^{j\psi(m,n)} = \sum_{{\beta}=1}^{2^{N_b-1}} c_{\beta} \prod_{p\in\mathcal{I}} s_{(m,n)}^{(p)},
\label{Nb-bit encoding}
\end{aligned}
\end{equation}
where $c_{\beta}$ denotes the encoding coefficient and $\mathcal{I}$ indexes the spin configurations participating in the $k^{th}$ interaction term. The phase encoding consists of $2^{N_b-1}$ product terms, each formed by a coefficient multiplied by a spin combination. These combinations are restricted to odd-order terms: $\binom{N_b}{1}$ first-order terms, $\binom{N_b}{3}$ third-order terms, $\binom{N_b}{5}$ fifth-order terms, etc., which collectively constitute the set $\mathcal{I}_k$. The encoding technique is illustrated through 1-bit, 2-bit, and 3-bit cases below.

1-bit encoding (\textit{$N_{b}$} = 1) is the simplest case, and the phase of each element can be directly expressed as one spin, as shown in Eq.~\ref{1bit-encoding}:
\begin{equation}
e^{j\psi(m,n)} = s_{(m,n)}^{(1)}.
\label{1bit-encoding}
\end{equation}
Here \textit{m} and \textit{n} stand for row index and column index of one antenna element, and $\psi{(m,n)}$ is the discrete phase of the element (either 0 or $\pi$).
Generally, the 1-bit encoding cannot provide sufficient degrees of freedom to deal with complex BF problems, 
and the 2-bit encoding was developed\cite{ross2021engineering}. 
In the 2-bit encoding (\textit{$N_{b}$} = 2), there are two spins ($s_{(m,n)}^{(1)}$ and $s_{(m,n)}^{(2)}$) in the phase expression as shown in Eq.~\ref{2bit-encoding}:
\begin{equation}
\begin{aligned}
    e^{j\psi(m,n)} = c_{1}s_{(m,n)}^{(1)} &+ c_{2}s_{(m,n)}^{(2)},
\end{aligned}
\label{2bit-encoding}
\end{equation}
where $c_{1}$ and $c_{2}$ are encoding coefficients and:
\begin{equation}
\textbf{\textit{S}}\textbf{\textit{c}} = \textbf{\textit{p}},\\
\label{phase-spin-relation}
\end{equation}
where $\textbf{\textit{S}}$ is a full-rank matrix extracted from all spin combinations, $\textbf{{\textit{c}}}$ is the coefficient vector, \textbf{\textit{p}} is the encoded phase vector. In 2-bit encoding, all possible spin combinations can be expressed by a $4\times2$ matrix: 
\begin{equation}
\begin{aligned}
\begin{bmatrix}
1 & -1 & -1 & 1\\
1 & 1 & -1 & -1 \\
\end{bmatrix}^T
\end{aligned}
\end{equation}
and all possible phase values are 0, $\frac{\pi}{2}$, $\pi$ and $\frac{3\pi}{2}$. To calculate $\textbf{\textit{c}} = \begin{bmatrix}{\textit{$c_{1}$}} \ {\textit{$c_{2}$}}\end{bmatrix}^{T}$, we extract the maximal linearly independent set as \textbf{\textit{S}}:
\begin{equation}
\textbf{\textit{S}} = \begin{bmatrix}
1 & 1\\1 & -1\\ \end{bmatrix}.
\end{equation}
Substituting $\textbf{\textit{p}}$ = $\begin{bmatrix}
e^{j0} \ e^{j\frac{\pi}{2}}\end{bmatrix}^{T}$ 
into Eq.~\ref{phase-spin-relation} we get:
\begin{equation}
\begin{bmatrix}
c_{1} \\c_{2}\end{bmatrix} = \textbf{\textit{S}}^{-1}\textbf{\textit{p}} = \begin{bmatrix}
\frac{1+j}{2} \\ \frac{1-j}{2}\end{bmatrix}.
\end{equation}

\begin{figure}[htbp]
\centering 
\includegraphics[width=0.6\textwidth]{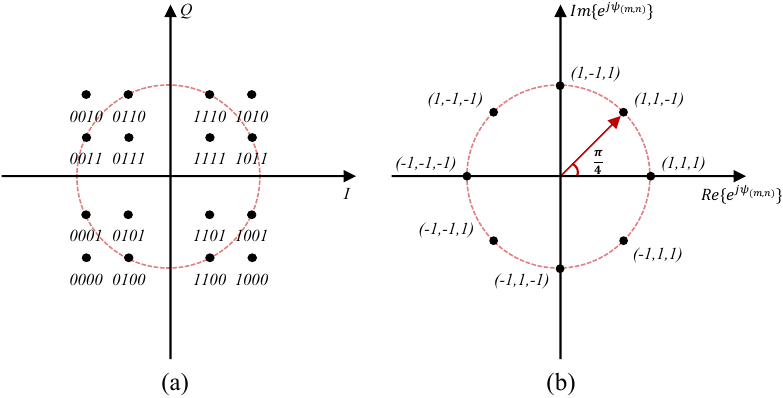}
\caption[The system.]{Phase mapping comparison between 16-QAM and our proposed encoding technique. (a): 16-QAM constellation. The circle drawn with a dashed line indicates the points with the same amplitude. (b): Proposed 3-bit encoding. Each point represents a spin combination ($s_{p}, s_{p+1}, s_{p+2}$).}
\label{16-QAM-vs-3-bit-encoding}
\end{figure}

When \textit{$N_{b}$} = 3, Eq.~\ref{Nb-bit encoding} under 3-bit encoding can be rewritten as:
\begin{equation}
\begin{aligned}
    e^{j\psi(m,n)} &= c_{1}s_{(m,n)}^{(1)} + c_{2}s_{(m,n)}^{(2)} + c_{3}s_{(m,n)}^{(3)} 
    \\&+ c_{4}s_{(m,n)}^{(1)}s_{(m,n)}^{(2)}s_{(m,n)}^{(3)},
\end{aligned}
\label{3-bit encoding}
\end{equation}
where coefficient vector $\textbf{\textit{c}} = \begin{bmatrix}
{\textit{$c_{1}$}}\
{\textit{$c_{2}$}}\
{\textit{$c_{3}$}}\
{\textit{$c_{4}$}}
\end{bmatrix}^{T}$ can be calculated from Eq.~\ref{phase-spin-relation} with the following \textbf{\textit{S}} and $\textbf{\textit{p}}$ :
\begin{equation}
\begin{aligned}
\textbf{\textit{S}} =
\begin{bmatrix}
1 & 1 & 1 & 1\\
1 & 1 &-1 & -1\\
1 & -1 & 1 & -1\\
1 & -1 & -1 & 1\\
\end{bmatrix},
\textbf{\textit{p}} = \begin{bmatrix}
e^{j0}\\
e^{j\frac{\pi}{4}}\\
e^{j\frac{\pi}{2}}\\
e^{j\frac{3\pi}{4}}
\end{bmatrix},\\
\end{aligned}
\label{3-bit-coding-coefficients}
\end{equation}
Here columns 1–4 of $\textbf{\textit{S}}$ map to $s_{(m,n)}^{(1)}$, $s_{(m,n)}^{(2)}$, $s_{(m,n)}^{(3)}$ and $s_{(m,n)}^{(1)}s_{(m,n)}^{(2)}s_{(m,n)}^{(3)}$ in Eq.~\ref{3-bit encoding}, respectively. For example, the first row stands for $s_{(m,n)}^{(1)}$ = 1, $s_{(m,n)}^{(2)}$ = 1, $s_{(m,n)}^{(3)}$ = 1 and $s_{(m,n)}^{(1)}s_{(m,n)}^{(2)}s_{(m,n)}^{(3)}$ = 1.
The total number of spin combinations should be $2^{3} = 8$, and we select a maximal linearly independent set to compose the full-rank matrix \textbf{\textit{S}}. As a result, Eq.~\ref{3-bit encoding} has a unique solution:
\begin{equation}
c_k = \frac{1}{4} \begin{cases}
\sqrt{4 + 2\sqrt{2}} \cdot e^{j\frac{3\pi}{8}}, & k=1 \\
\sqrt{4 + 2\sqrt{2}} \cdot e^{-j\frac{\pi}{8}}, & k=2 \\
\sqrt{4 - 2\sqrt{2}} \cdot e^{-j\frac{\pi}{8}}, & k=3 \\
\sqrt{4 - 2\sqrt{2}} \cdot e^{-j\frac{5\pi}{8}}, & k=4.
\end{cases}
\end{equation}
The proposed encoding technique achieves higher efficiency in phase-only optimization compared with conventional encoding techniques. For example, 16-QAM\cite{dupuis197916,chong2003new,kim2019leveraging} is a widely used encoding technique with 4 binary variables employed to compose a constellation including 16 points in the complex plane (shown in Fig. \ref{16-QAM-vs-3-bit-encoding}(a)). 
One can cover at most 8 points sharing the same amplitude in Fig. \ref{16-QAM-vs-3-bit-encoding}(a); thus, 16-QAM suffers from low efficiency in the phase-only encoding. While from Fig. \ref{16-QAM-vs-3-bit-encoding}(b), it is clear that every spin combination is mapped onto a point at the unit circle, and we achieve phase-only encoding with fewer variables. 
For higher-precision encoding techniques (4-bit or higher), we can construct a full-rank matrix \textbf{\textit{S}} and determine the corresponding encoding coefficients using the method mentioned before. Therefore, our method remains applicable in even more complex scenarios requiring high-precision encoding.

\subsection{Beamforming Hamiltonian}
Using dyadic Green's function, we could obtain a far-field radiation pattern of a single element with a size of $d\times d$ based on the source current $\mathbf{J}_{\mathbf{s}}$\cite{balanis2016antenna,ross2021engineering}, see Eq.~\ref{epsilon}:
\begin{equation}
\begin{aligned}
&\mathbf{E} = - j\omega\mu\frac{e^{- jkr}}{4\pi r}\left( {I + \frac{1}{k^{2}}\nabla\nabla} \right){\int{\mathbf{J}_{\mathbf{s}}\left( \mathbf{r'} \right)e^{jk{\hat{\mathbf{r}} \cdot \mathbf{r'}}d\mathbf{r'}}}},\\
&E_{\theta} = - \frac{jkd^{2}}{2\pi r}e^{- jkr}{\cos\theta}\ {\cos\phi}\ \rm{sincX}\  \rm{sincZ},\\
&E_{\phi} = \frac{jkd^{2}}{2\pi r}e^{- jkr}{\sin\phi}\ \rm{sincX}\  \rm{sincZ},\\
&\rm{X} = \frac{\textit{kd}\ {\sin\theta}\ {\cos\phi}}{2},\ \rm{Z} = \frac{\textit{kd}\ {\cos\theta}}{2}.
\end{aligned}
\label{epsilon}
\end{equation}
Here $k$ represents the wave number, $\omega$ is the angular frequency and $\mu$ is the permeability. In BF optimization, the objective function is the total radiated power of the array in the desired direction. For an \textit{M} $\times$ \textit{N}  array, the array factor can be written as Eq.~\ref{AF}.
%
\begin{equation}
\begin{aligned}
AF &= {\sum\limits_{m = 1}^{M}{\sum\limits_{n = 1}^{N}{e^{j{({k_{x}md + k_{z}nd})}}e^{j\psi(m,n)}}}},\\
k_{x} &= k\ {\sin\theta}\ {\cos\phi},\ k_{z} = k\ {\cos\theta}.
\end{aligned}
\label{AF}
\end{equation}

The BF Hamiltonian \textit{$H_{BF}$} is expressed by the negative of total radiated power within the desired region (solid angle integration centred at the target angle defined by $\theta_{BF}$ and $\phi_{BF}$) as is shown in Eq.~\ref{Array synthesis}.
\begin{equation}
\begin{aligned}
H_{BF} &= - {\iint{\left| \mathbf{E} \right|^{2}\left| {AF} \right|^{2}}}d\mathbf{\Omega},\\
{\iint{d\mathbf{\Omega}}} &= {\int_{\phi_{BF} - \frac{\Delta\phi}{2}}^{\phi_{BF} + \frac{\Delta\phi}{2}}{d\phi{\int_{\theta_{BF} - \frac{\Delta\theta}{2}}^{\theta_{BF} + \frac{\Delta\theta}{2}}{{\sin\theta}d\theta}}}}. 
\end{aligned}
\label{Array synthesis}
\end{equation}
Substituting the spin representation of phases into the array radiation pattern yields the \textit{$H_{BF}$} with \textit{${N_{b}}$}-bit phase encoding:
\begin{equation}
\begin{aligned}
H_{BF}=\sum\limits_{m}^M\sum\limits_{n}^N\sum\limits_{\substack{i}}^M\sum\limits_{\substack{v}}^NJ_{(m,n,i,v)}^{(p,q)}\prod_{p \in \mathcal{I}} s_{(m,n)}^{(p)} \prod_{q \in \mathcal{I}} s_{(i,v)}^{(q)},\\
\end{aligned}
\label{K-bit HBF}
\end{equation}
where $J_{(m,n,i,v)}^{(p,q)}$ indicates the spin-spin interaction between element $(m, n)$ and element $(i, v)$, and $\mathcal{I}$ is the set mentioned in Eq.~\ref{Nb-bit encoding}. $J_{(m,n,i,v)}^{(p,q)}$ can be calculated by:
\begin{equation}
\begin{aligned}
J_{(m,n,i,v)}^{(p,q)} &= - c_{\beta}c_{{\beta}'}^{*}{\iint{\left| \mathbf{E} \right|^{2}e^{j{\lbrack{k_{x}{({m - i})}d + k_{z}{({n - v})}d}\rbrack}}d\mathbf{\Omega}}},\\
\end{aligned}
\label{spin-spin interaction}
\end{equation}
where the array size is ${M}\times{N}$ with an element spacing of $d$, $c_{\beta}$ and $c_{{\beta}'}^{*}$ are the coefficients defined in Eq.~\ref{Nb-bit encoding}. The number of spin-spin interactions depends on the number of terms in phase expression and the array size.
In 1-bit encoding, the phase of each element can be expressed by a single spin, thus the interaction between two elements contains only  one term and the total number of $J_{(m,n,i,v)}^{(p,q)}$ is $\binom{MN}{2}$. While in 2-bit encoding, the phase expression is composed of two terms and the interaction contains four terms and the total number of $J_{(m,n,i,v)}^{(p,q)}$ is $4\times \binom{MN}{2}$. In 3-bit phase encoding, high-order terms transform the optimization into a HUBO problem. The phase expression comprises four terms, including three first-order terms and a third-order term. Interactions between unit pairs involve sixteen terms: two first-order terms generate a second-order term, a first-order and third-order term generate a fourth-order term, and two third-order terms generate a sixth-order term, resulting in a total of $16 \times \binom{MN}{2}$ $J_{(m,n,i,v)}^{(p,q)}$. In more general cases, when employing $N_{b}$-bit phase encoding, there exists $4^{N_{b}-1}\times \binom{MN}{2}$ $J_{(m,n,i,v)}^{(p,q)}$.

Furthermore, the Hamiltonian computation can be significantly accelerated through parallel processing at two levels. First, in multi-objective optimization scenarios, the independent Hamiltonians corresponding to different BF directions or null regions can be computed concurrently. Second, within each Hamiltonian configuration, the spin-spin interactions are fully decoupled and can be efficiently calculated through matrix-based parallel computation, leading to substantial reductions in Hamiltonian computation time. This parallelization strategy is particularly effective for large-scale array configurations, where it achieves near-linear speedup while maintaining computational accuracy.
Precomputed Hamiltonians could be stored and reused, enabling real-time adaptation to target direction ($\theta_{BF},\phi_{BF}$) without recalculation.

\subsection{SB Optimizer}  
The Simulated Bifurcation (SB) optimizer, proposed by Goto et al. \cite{goto2016bifurcation}, is a powerful algorithm for solving large-scale combinatorial optimization problems. It simulates the adiabatic evolution of nonlinear Hamiltonian systems, where bifurcation phenomena map to spin states (+1 or -1). The SB optimizer is particularly suited for parallel computing due to its simultaneous updating mechanism \cite{goto2019combinatorial}.

The corresponding classical Hamiltonian of the SB optimizer, which is derived from a classical approximation of the quantum Hamiltonian, is defined as:  
\begin{equation}
H_c(\mathbf{x},\mathbf{y},t) = \sum_{i = 1}^{N_s}\left[\frac{K}{4}(x_i^2+y_i^2)^2-\frac{p(t)}{2}(x_i^2-y_i^2)+\frac{\Delta_i}{2}(x_i^2+y_i^2)\right] - \frac{\xi_0}{2} H_{BF}(t),
\label{classical Hamiltonian}
\end{equation}
where $x_i$ and $y_i$ are a pair of canonical conjugate variables for the $i^{th}$ Kerr-nonlinear parametric oscillator. $N_{s}$ = ${M}\times{N}\times{N_{b}}$, represents the total spin number. The term $K$ stands for the positive Kerr coefficient. The expression $p(t)$ describes the time-dependent amplitude of parametric two-photon pumping. The variable $\Delta_i$ indicates the positive detuning frequency, which is the difference between the resonance frequency of the $i^{th}$ oscillator and half of the pumping frequency. $\xi_0$ is a positive constant that plays a vital role in evolution. Moreover, $H_{BF}(t)$ is the objective Hamiltonian of the BF optimization problem as defined in Eq. \ref{K-bit HBF}, substituting $E_{Ising}$ mentioned in \cite{goto2019combinatorial}. However, we set it time-dependent here because it may change with time according to the application scenarios, which will be illustrated in 3.1.1. Initially, each oscillator is in a vacuum state, and $\xi_0$ is chosen small enough to ensure the vacuum state is the ground state of the initial Hamiltonian. As the pumping strength $p(t)$ gradually increases, each oscillator transitions to a coherent state with a positive or negative amplitude. According to the adiabatic evolution theorem, the system remains in its ground state throughout this process, and the final amplitude sign corresponds to the spin in the ground state of $H_{BF}(t)$.  

While the general framework of SB provides a powerful approach to combinatorial optimization, we choose the ballistic simulated bifurcation(bSB) algorithm in this paper \cite{zeng2024performance, goto2021high}. The core of bSB lies in its modified equations of motion and the incorporation of boundary conditions that emulate a particle's movement and interaction with the environment. The equations of motion for this classical mechanical Hamiltonian for fast numerical simulation are derived by disregarding some terms proportional to $\mathbf{y}$. The simplified Hamiltonian $H_{\mathrm{SB}}(\mathbf{x}, \mathbf{y}, t)$ is represented as \begin{equation}
\dot{x}_i = \frac{\partial H_{\mathrm{SB}}}{\partial y_i} = \Delta y_i,
\label{SB Hamiltonian for y}
\end{equation}
\begin{equation}
\dot{y}_i = -\frac{\partial H_{\mathrm{SB}}}{\partial x_i} = -\left[ \Delta - p(t) \right] x_i - \frac{\xi_{0}}{2}\frac{\partial H_{BF}(t)}{\partial x_{i}}.
\label{bSB Hamiltonian for x}
\end{equation}
where $\Delta$ is the unified detuning frequency. These equations are solved using the explicit fourth-order Runge-Kutta method \cite{gottlieb1998total}. All variables $\mathbf{x}$ and $\mathbf{y}$ are initialized around zero. As $p(t)$ increases, each oscillator transitions to a coherent state, and the sign of the final $x_i$ determines the $i^{th}$ spin state. The optimized phase configuration of the MIMO array is obtained by decoding the final spin states using the corresponding encoding technique. After each update of $x_i$, if its absolute value exceeds a predefined threshold (typically 1), $x_i$ is clipped to $\text{sgn}(x_i)$, corresponding to spin $s_i$ as explained in Eq.~\ref{Ising Hamiltonian}. And its corresponding momentum $y_i$ is reset to zero. This mechanism simulates a collision with a boundary, causing the particle to lose its momentum and essentially bounce back, preventing it from straying too far and facilitating escape from shallow local minima. 

To clarify, in our study, the MIMO beamforming problem becomes a QUBO problem with 1-bit or 2-bit phase encoding where spin interactions are limited to quadratic terms. However, when using 3-bit or higher precision encoding (e.g., 4-bit), higher-order terms emerge, transforming the problem into a HUBO formulation. Conventional quantum annealing approaches typically introduce auxiliary variables to reduce the order of interactions\cite{boros2002pseudo,jiang2018quantum,mato2022quantum}, converting higher-order terms to quadratic ones.  In contrast, our proposed method with 3-bit or higher precision encoding maintains the higher-order terms in $H_{BF}(t)$ in Eq. \ref{bSB Hamiltonian for x}, which do not affect the ordinary differential equation-solving process. SB algorithm directly handles higher-order interactions without requiring auxiliary variables\cite{kanao2022simulated}, thereby effectively reducing computational complexity.

The SB optimizer offers significant advantages, including the ability to simulate quantum-like behavior on classical computers and support for massively parallel processing.

\section{Results and discussions}
\subsection{Experiment Setting}
\subsubsection{Application Scenarios}
This section presents three applications of our quantum-inspired algorithm on a 10$\times$24 antenna array (240 elements): single beam, multi-beam with nulls and single beam with sidelobe suppression. Using 1, 2, and 3-bit encoding, we map the quantized phases to spins and compute spin-spin interactions. The SB algorithm determines the (near) ground state spin configuration.

\paragraph{\textbf{Single Beam:}}

We aim to realize a single BF at the angular coordinates $\theta = 50^{\circ}$ and $\phi = 50^{\circ}$. For an array configured in the $xoz$ plane, this constitutes a significant steering angle. Within the framework of the SB algorithm, the Hamiltonian $H_{BF}$ is explicitly defined in Eq.~\ref{K-bit HBF}. It is worth noting that here the original spin variable $s$(Eq.~\ref{K-bit HBF}) should be replaced by its continuous counterpart $x$(Eq.~\ref{classical Hamiltonian}), which is discretized back to +1 or -1 after the solution is obtained. The construction of the subsequent Hamiltonian follows this principle.

\paragraph{\textbf{Multiple Beams with Nulls:}}

The capability of generating multiple beams with nulls between them is essential to avoid interference and improve throughput in communication. 
we aim to realize BF at ($\theta$ = $60^{\circ}$, $\phi$ = $75^{\circ}$), ($\theta$ = $90^{\circ}$, $\phi$ = $75^{\circ}$) and ($\theta$ = $120^{\circ}$, $\phi$ = $75^{\circ}$), and nulls at ($\theta$ = $75^{\circ}$, $\phi$ = $75^{\circ}$) and ($\theta$ = $105^{\circ}$, $\phi$ = $75^{\circ}$).
Clearly, this scenario inherently involves multi-objective optimization (shown in Eq.~\ref{multi-obj-cost-cuntion2}), and we balance the power distribution of these beams by assigning different weights to them based on their energy contributions.
\begin{equation}
\begin{aligned}
&H_{BF}(t) = H_{beams}(t)+H_{nulls}(t),\\
&H_{beams}(t) = {\sum\limits_{o = 1}^{N_{beams}}{w_{o}(t)H}_{BF,o}} = - \sum\limits_{o=1}^{N_{beams}}w_{o}(t){\iint{\left| \mathbf{E} \right|^{2}\left| {AF} \right|^{2}}}d\mathbf{\Omega}_{o},
\\
&H_{nulls}(t) = {\sum\limits_{i = 1}^{N_{nulls}}{w_{i}(t)H}_{null,i}} = {\sum\limits_{i = 1}^{N_{nulls}}{w_{i}(t){\iint{\left| \mathbf{E} \right|^{2}\left| {AF} \right|^{2}}}d\mathbf{\Omega}_{i}}},\\
&{\iint{d\mathbf{\Omega}_{o,i}}} = {\int_{\phi_{o,i} - \frac{\Delta\phi}{2}}^{\phi_{o,i} + \frac{\Delta\phi}{2}}{d\phi{\int_{\theta_{o,i} - \frac{\Delta\theta}{2}}^{\theta_{o,i} + \frac{\Delta\theta}{2}}{{\sin\theta}d\theta}}}},
\label{multi-obj-cost-cuntion2}
\end{aligned}
\end{equation}
where $H_{beams}(t)$ is multi-BF Hamiltonian (the number of beams is $N_{beams}$) and $H_{nulls}(t)$ is multi-nulls Hamiltonian (the number of beams is $N_{nulls}$). We use a positive sign for each term in $H_{nulls}(t)$ because lower sidelobe levels correspond to lower energy in the objective function. $w_{o}(t)$ and $w_{i}(t)$ are the weights of the BF Hamiltonian $H_{BF,o}$ and null Hamiltonian $H_{null,i}$, respectively. These weights could be seen as a penalty-based approach employed in these multi-objective combinatorial optimizations. In this case, they balance each beam's contribution, helping force a relatively equal energy distribution across beams. Specifically, we employ a 1/cos($\theta_{BF,o}$) weighting scheme originally and adjust it with time, which compensates for the natural angular variation in array gain patterns. This time-varying weight adjustment ensures the balanced evolution of the target Hamiltonian.

\paragraph{\textbf{Single Beam with Wide Range Upper Sidelobe Suppression:}}
In multiple receivers or transmitters scenarios, the BF problem is a complex combinatorial optimization that requires achieving multiple objectives: BF in the target area and suppressing sidelobes in non-target areas. 
We aim to realize single BF at $\theta$ = $90^{\circ}$ with sidelobe suppression from $\theta$ = $0^{\circ}$ to $85^{\circ}$, 
Within the framework of the SB algorithm, the target Hamiltonian, including a single beam and many sidelobes, is represented in the same form as Eq.~\ref{multi-obj-cost-cuntion2}. Similarly, $w_{o}(t)$ and $w_{i}(t)$ are the weights of the BF Hamiltonian $H_{BF,o}$ and sidelobe suppression Hamiltonian $H_{null,i}$, respectively. The large region of sidelobe suppression is divided into $N_{nulls}$ sub-regions centered at ($\theta_{i}$, $\phi_{i}$). We assign a weight of 10 to $H_{BF,o}$ ($w_{o}(0)$)and 1 to $H_{null,i}$ ($w_{i}(0)$) originally, and change the weights with time. This penalty-based approach ensures solutions achieve both optimal BF and effective sidelobe suppression.

\subsubsection{Compared Benchmarks}
For comparison, we conducted experiments on a single-core Intel(R) Core(TM) i7-9700 processor and coding language Python.
To evaluate the performance of our proposed method, we compare it against several benchmark algorithms: semi-analytical holographic BF, digital BF with CMV, genetic algorithm (GA), and simulated annealing (SA). Below, we briefly describe the principles and parameter settings for each benchmark method.

\paragraph{\textbf{Holographic Beamforming:}}
The holographic BF algorithm aims to configure the phase of each array element to radiate toward a pre-defined BF direction\cite{black2017holographic}. The phase delay $\phi(x, z)$ of an element located at $(x, z)$ is derived from the superposition of multiple beams, as Eq. \ref{HBF}:
\begin{equation}
\phi(x, z)=\arg \left\{\sum_{l=1}^{N_{beams}} A_l e^{jk_l\left(x \cos \theta_l \cos \phi_l+z \sin \phi_l\right)} \right\}.
\label{HBF}
\end{equation}
Here, $A_l$ is the amplitude of the $l^{th}$ beam, and $(\theta_l, \phi_l)$ represents the BF direction. The solution is continuous but is quantized to discrete values for fair comparison with SB. 

\paragraph{\textbf{Digital Beamforming with Constrained Minimum Variance (CMV):}}
 
DBF precisely controls the phase and amplitude of each antenna. The array factor $AF(\theta, \phi)$ is expressed as Eq.~\ref{AF}. The far field is represented in matrix form as $\mathbf{A x} = \mathbf{b}$, where $\mathbf{A}$ is the array factor matrix, $\mathbf{x}$ is the excitation vector, and $\mathbf{b}$ is the electric field vector.
The CMV optimization minimizes the output power while maintaining a unit gain in the BF direction. The objective function and constraint are:
\begin{equation}
\min _{\mathbf{x}} \mathbf{x}^H \mathbf{A}^H \mathbf{A} \mathbf{x},  \text{subject to} \quad \mathbf{x}^H \mathbf{A}\left(\theta_{BF}, \phi_{BF}\right)=1.
\end{equation}

\begin{figure}[htbp]
\centering 
\includegraphics[height = 0.45\textwidth]{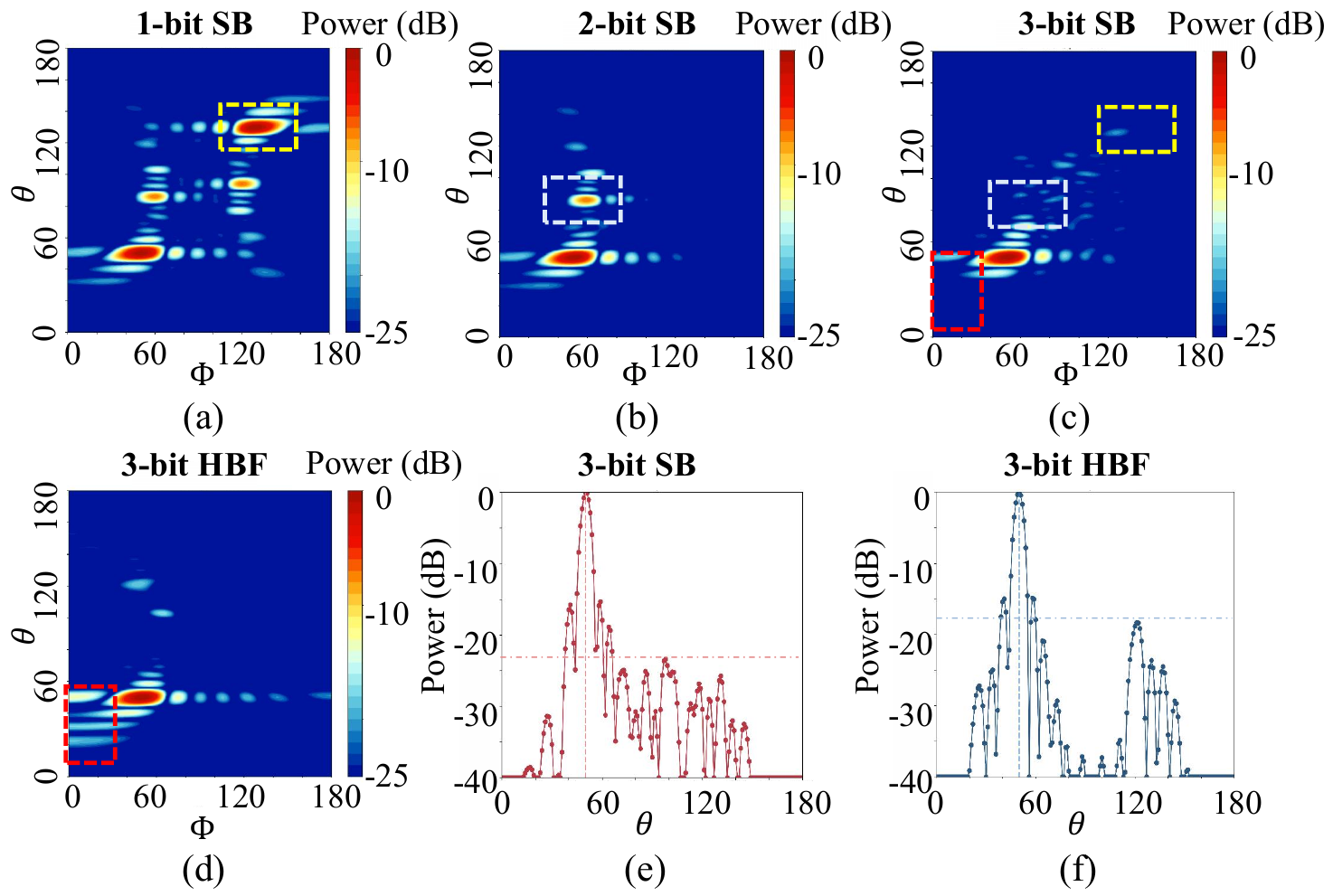}
\caption[The system.]{Single BF at $\theta$ = $50^{\circ}$, $\phi$ = $50^{\circ}$ of $10\times24$ array with different encoding precision by SB and holographic BF (HBF) algorithms. (a)-(c): 2D far-field patterns by SB with 1, 2 and 3-bit encoding. (d): 2D far-field patterns by the holographic algorithm with 3-bit encoding. (e)-(f): 1D ($\phi$ = $50^{\circ}$) far-field pattern by SB and HBF with 3-bit encoding.}
\label{single beam}
\end{figure} 

The optimal solution is derived using the Lagrange multiplier method, yielding:
\begin{equation}
\mathbf{x}_{opt}=\frac{\mathbf{A}^H \mathbf{A}^{-1} \mathbf{A}\left(\theta_{BF}, \phi_{BF}\right)}{\mathbf{A}^H\left(\theta_{BF}, \phi_{BF}\right) \mathbf{A}^H \mathbf{A}^{-1} \mathbf{A}\left(\theta_{BF}, \phi_{BF}\right)}.
\end{equation}
For phase-only arrays, the magnitude is neglected, and the phase is quantized to the nearest discrete value.

\paragraph{\textbf{Genetic Algorithm (GA):}}
 
GA is an evolutionary optimization method that mimics natural selection\cite{Mirjalili2019}. It iteratively evolves a population of candidate solutions through selection, crossover, and mutation operations to converge toward an optimal solution. In this section, GA is employed as a benchmark in the single beam with upper sidelobe suppression scenario. The algorithm initializes a population of 150 individuals with 3-bit quantized phase values and evolves over 300 generations. Tournament selection with a size of 3, uniform crossover, and adaptive mutation (base rate of 0.08, increasing for lower fitness) ensure diverse and high-quality offspring.
The fitness function balances the target gain and sidelobe suppression, as shown in Eq. \ref{GA},

\begin{equation}
f(\mathbf{w}) = BF - \sum\limits_{i}^{N_{nulls}} \alpha_i \cdot Null_i,
\label{GA}
\end{equation}
where \(\mathbf{w}\) is the weight vector representing the phase configuration of the antenna array,
\(BF = \left| \mathbf{w}^H \mathbf{A}(\theta_{BF}, \phi_{BF}) \right|^2\) represents the gain in the BF direction,
\(Null_i = \left| \mathbf{w}^H \mathbf{A}(\theta_i, \phi_i) \right|^2\) represents the gain across the $i^{th}$ sidelobe direction,
and \(\alpha_i\) is a trade-off parameter controlling the $i^{th}$ sidelobe suppression. This formulation ensures the GA optimizes for both high directivity in the target direction and sidelobe suppression.

\begin{figure}[htbp]
\centering 
\includegraphics[height=0.45\textwidth]{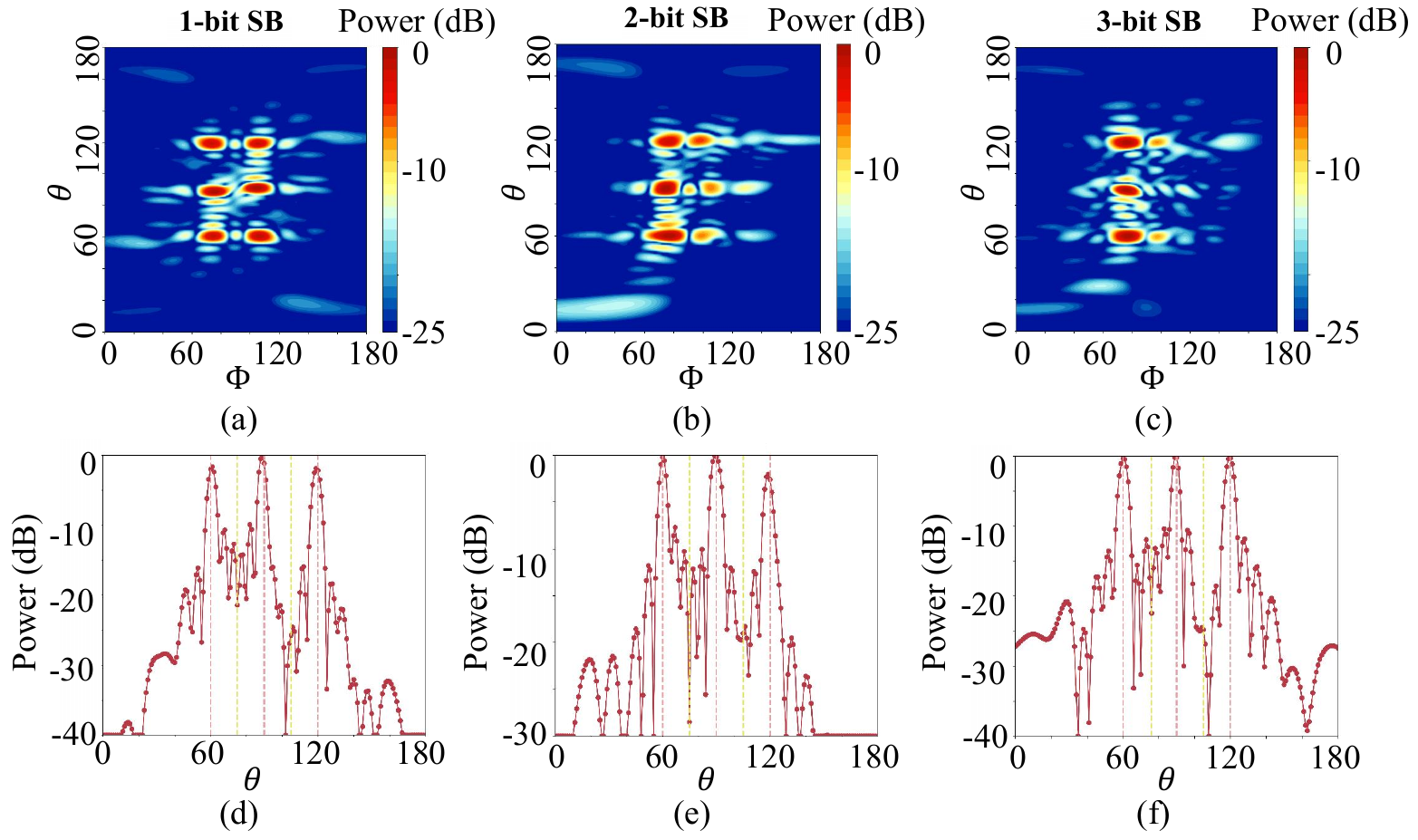}
\caption[The system.]{Multiple beams with nulls optimization of $10\times24$ array by SB with different encoding precision. BF at ($\theta$ = $60^{\circ}$, $\phi$ = $75^{\circ}$), ($\theta$ = $90^{\circ}$, $\phi$ = $75^{\circ}$) and ($\theta$ = $120^{\circ}$, $\phi$ = $75^{\circ}$). Nulls are located at ($\theta$ = $75^{\circ}$, $\phi$ = $75^{\circ}$) and ($\theta$ = $105^{\circ}$, $\phi$ = $75^{\circ}$). (a)-(c): 2D far-field pattern of 1, 2 and 3-bit encoding by SB. (d)-(f): 1D ($\phi$ = $75^{\circ}$) far-field pattern of 1, 2 and 3-bit encoding by SB.}
\label{3beams}
\end{figure} 

\paragraph{\textbf{Simulated Annealing (SA):}}

SA is a probabilistic optimization scheme inspired by the annealing process in metallurgy\cite{Vecchi_SA}. It explores the solution space by accepting suboptimal solutions with decreasing probability, ultimately converging to a global optimum. SA is employed as a benchmark for optimizing the phase configurations in the single beam with upper sidelobe suppression scenario, utilizing a 3-bit quantized phases to ensure practical feasibility. The algorithm starts with an initial temperature of 1,000 and iteratively cools down at a rate of 0.95 over 4,000 iterations, accepting suboptimal solutions with a probability determined by the current temperature and the difference in objective values. The fitness function balances the target gain and sidelobe suppression with a trade-off parameter $\alpha_i$ the same as Eq. \ref{GA}. Multiple runs are performed to ensure robustness.

These benchmarks provide a comprehensive basis for evaluating the performance of proposed BF approaches, ensuring a robust comparison across different strategies.


\subsection{Comparison Results}
The results are presented in order of increasing optimization complexity, ranging from simple scenarios to more challenging ones. We compare the performance of different algorithms focusing on optimization quality and computational efficiency. Additionally, we explore the performance and limitations of different bit encodings, evaluating their impact on BF flexibility and precision in various scenarios. This analysis provides insights into the trade-offs between phase quantization levels and optimization outcomes in complex BF tasks.
\subsubsection{Single Beam}
Holographic BF is commonly used for simple objectives such as single beam. We compare it with our method using different bit encodings (1-bit, 2-bit, and 3-bit), as shown in Fig. \ref{single beam}. As illustrated in Fig. \ref{single beam}(a)-(c), with the encoding precision increases, the radiated power becomes more concentrated in the desired direction. Due to the low phase precision of 1-bit and 2-bit encoding, BF flexibility is limited, resulting in inevitable high sidelobes. The sidelobes are significantly reduced with 3-bit encoding.

Comparing Fig. \ref{single beam}(c) with Fig. \ref{single beam}(d), particularly in the region marked with the red dashed line and the 1D profiles in Fig. \ref{single beam}(e) and (f), it is evident that while holographic BF provides near-instantaneous solutions, our method exhibits less energy leakage, lower sidelobes, and reduced interference clutter.


\subsubsection{Multiple Beams with Nulls}
{To further investigate the limitations of different bit encodings in complex scenarios, we apply the SB algorithm with 1, 2, and 3-bit encoding to optimize the proposed three-beam, two-null configuration, as shown in Fig. \ref{3beams}. It is evident that SB with 1-bit and 2-bit encoding fails to suppress unwanted beams and struggles to achieve three energy-balanced beams, with their outputs consistently exhibiting symmetric patterns due to the constraints of low phase precision. In contrast, the proposed 3-bit encoding enables SB-based optimization to generate three distinct beams and achieve nulls below -20 dB, demonstrating significantly improved performance in complex BF tasks.


\begin{figure*}[htbp]
\centering 
\includegraphics[width=1\textwidth]{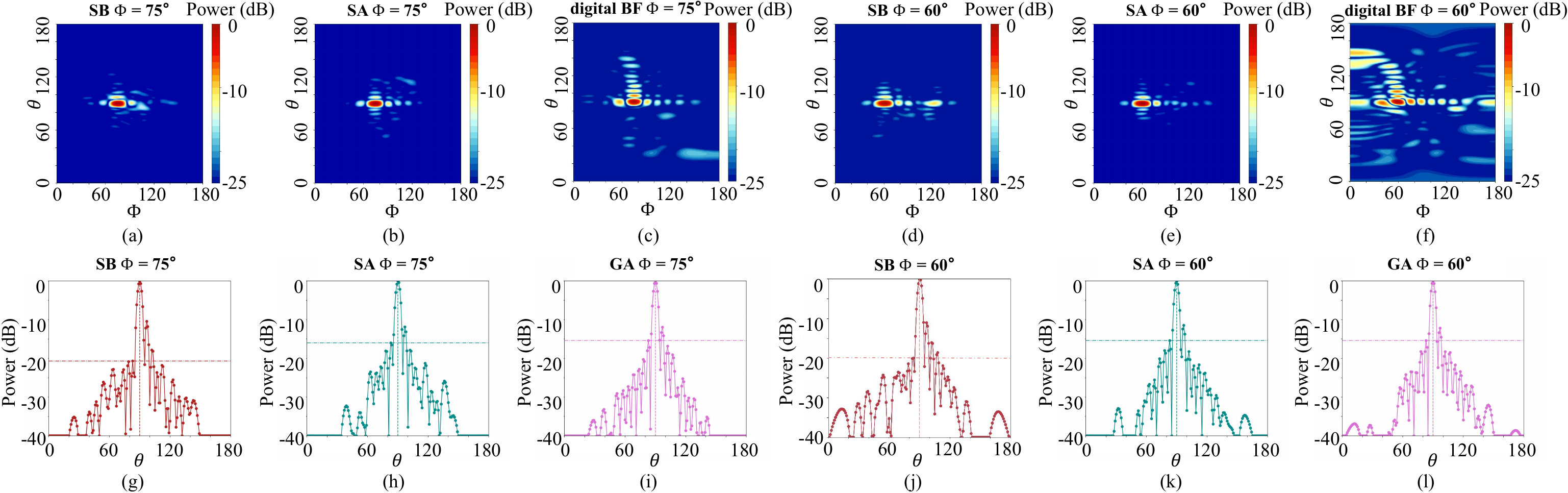}
\caption[The system.]{Single BF with sidelobe suppression from $\theta$ = $0^{\circ}$ to $85^{\circ}$ of $10\times24$ array by SB, SA, GA and digital BF in 3-bit encoding. (a)-(f): 2D plot of far-field pattern optimized by SB, SA and digital BF. (g)-(l): 1D plot of far-field pattern optimized by SB, SA and GA. BF at $\theta$ = $90^{\circ}$, $\phi$ = $75^{\circ}$ ((a)-(c), (g)-(i)) and $\theta$ = $90^{\circ}$, $\phi$ = $60^{\circ}$ ((d)-(f), (j)-(l)).}
\label{sidelobe}
\end{figure*} 

\begin{figure*}[htbp]
\centering 
\includegraphics[width=0.92\textwidth]{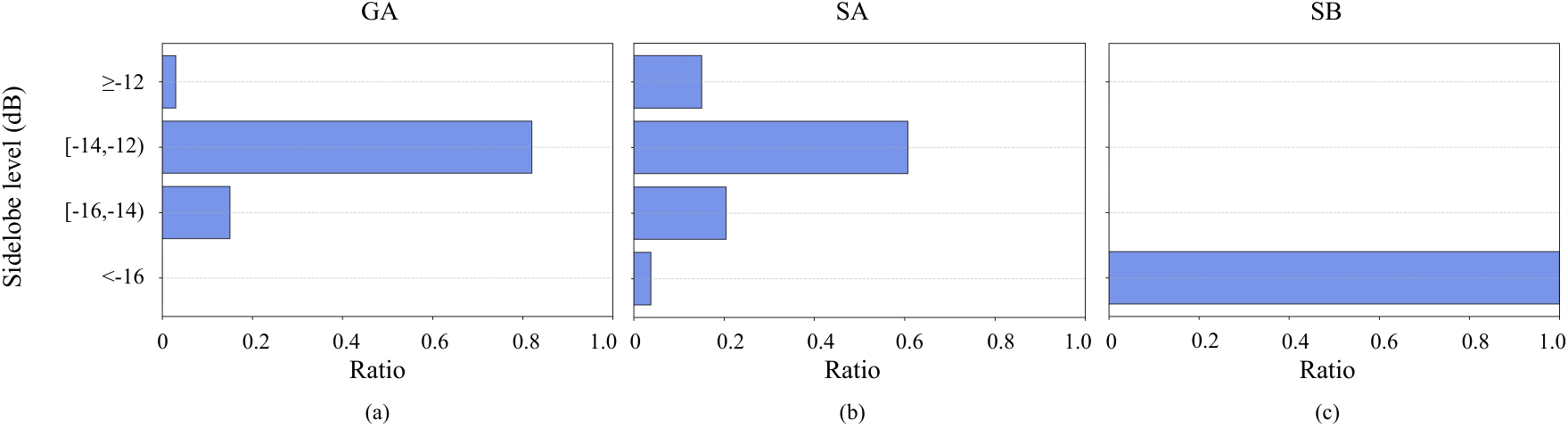}
\caption[The system.]{Sidelobe level distribution. Statistics of the optimized solutions for GA (a), SA (b) and SB (c) in the single BF ($\theta=90^\circ, \phi = 75^\circ$) with upper sidelobe suppression scenario. The sidelobe levels are statistically analyzed after running each of the optimization algorithms 1,000 times independently.}
\label{SLL_histogram}
\end{figure*} 

\subsubsection{Single Beam with Upper Sidelobe Suppression}

For this complex scenario, holographic BF is no longer suitable. Instead, we utilize digital BF, GA, and SA to evaluate their performance in comparison with our proposed SB algorithm.
Given the stringent requirements for wide-range sidelobe suppression, which exceed the capabilities of 1-bit and 2-bit techniques, we adopt 3-bit encoding. The BF results are illustrated in Fig. \ref{sidelobe}. 
For quantitative analysis, we focus on the example of BF at \((\theta = 90^\circ, \phi = 75^\circ)\), as detailed in the following sections.

\paragraph{\textbf{Sidelobe level:}} SB achieves significant optimization, particularly in the wide suppression angular range, where the sidelobe level is below -20 dB as shown in Fig. \ref{sidelobe}(a), (d). Among them, SB demonstrates superior performance in sidelobe suppression compared to GA (-15.39 dB/ -15.39 dB) and SA (-16.3 dB/ -15.39 dB). While conventional digital BF achieves comparable sidelobe suppression levels, this approach merely displaces the energy randomly throughout the pattern as detailed in its suboptimal directivity. In contrast, SB maintains controlled energy redistribution while achieving superior, spatially consistent suppression.

\paragraph{\textbf{Directivity:}} All optimization algorithms surpass conventional digital BF (565.81) when BF at $\theta$ = $90^{\circ}$, $\phi$ = $75^{\circ}$: GA achieves 705.32, SA attains 586.67, and SB yields 572.12. While SB’s directivity is lower, this results from its rigorous sidelobe suppression: energy displaced from the upper sidelobe region redistributes across the other side. This behavior is not unique to SB; if further increasing the weight factor $\alpha_i$ similarly in Eq. \ref{GA}, we found that the directivity by GA/SA will dramatically degrade and chaotic patterns are to be generated. Differently, SB shows clean and regular patterns with high level of suppression even with a large $\omega_i$ in Eq. \ref{multi-obj-cost-cuntion2}. Thus, while directivity often serves as a benchmark, SB’s superior upper-plane energy cleanliness and effective concentration in BF direction align with our primary objective for interference-sensitive applications.

\paragraph{\textbf{Time-to-Target (TTT):}}
The Time-to-Target (TTT) is a key metric for evaluating the computational efficiency of optimization algorithms. It is calculated as:
\begin{equation}
TTT = \frac{\ln(1 - P_d)}{\ln(1 - P_s(N))} \times t,
\label{TTT}
\end{equation}
where \( P_d \) is the desired success probability (set to 99\% for high reliability), \( P_s(N) \) is the success probability of finding the target solution in a single run, and \( t \) is the time for one run. It quantifies the expected computation time required for an optimization algorithm to reach the target solution with a specified confidence level. We conducted 1,000 independent experiments to ensure statistical significance.
The search space, defined as \(2^{N_{b} \times M \times N}\), is prohibitively large, making brute force methods infeasible for determining the ground state. 
To establish performance benchmarks, we first define a target solution criterion. The unsuppressed beam pattern exhibits a sidelobe level of $-12.3$\,dB, leading us to set $-15.3$\,dB (half-power reduction) as the suppression requirement for the upper sidelobe region. All timing metrics for the SB algorithm incorporate both Hamiltonian computation and solver execution phases.

For comparative evaluation, we adopt distinct methodologies for each algorithm. The SA and GA benchmarks employ fixed optimal parameters (determined through preliminary tuning) with 1,000 randomized initializations. In contrast, the SB evaluation simulates real-world deployment by testing 1,000 logarithmically spaced $\xi_0$ values between $10^{-8}$ and $1$, with each parameter set executed using random initial conditions. This strategy capitalizes on parallel parameter tuning, ensuring nearly identical optimization time whether processing a single parameter set or 1,000 configurations, as demonstrated in the following section. This approach provides comprehensive statistical analysis while maintaining practical relevance for actual implementation scenarios.

The optimization results shown in Fig. \ref{SLL_histogram} from 1,000 trials reveal distinct performance characteristics: GA achieved sidelobe levels of -13.5 ± 1.8 dB, but with merely 0.9$\%$ probability to meet the -15.3 dB threshold; SA showed higher variance (-13.1±6.1 dB), yielding 7$\%$ success probability but with 15$\%$ risk of degradation above -12 dB; notably, SB achieved 100$\%$ success probability in all trials with hundreds valid solutions per parameter sweep (in one trial). Furthermore, SB could achieve solutions with sidelobe suppression below -20 dB, a performance level unattainable by either our GA or SA implementations. A comprehensive performance comparison is presented in TABLE \ref{tab:performance}, which documents the optimal results obtained in our experimental evaluations.
Our proposed SB achieves a TTT of 55 s, outperforming GA 692,756 s and SA 13,960 s. While the average single-run time of SA and GA is 5 and 30 times higher than that of SB, the TTT of SB is significantly lower, indicating that SB achieves a higher success probability in finding the target solution. This demonstrates the SB algorithm's superior efficiency in finding optimal solutions while maintaining high reliability, highlighting the advantage of SB in reducing the overall computational overhead for complex optimization tasks. Additionally, while digital BF achieves satisfactory sidelobe suppression when $\phi$ = 75 $^\circ$, it exhibits two critical limitations: irregular high-energy leakage persists across the entire upper plane, and severe performance degradation occurs in other specific scenarios (e.g., at $\theta = 90^\circ$, $\phi = 60^\circ$), as shown in Fig. \ref{sidelobe}(f). This fundamentally restricts its practical applicability in real-world deployment scenarios.

\begin{table}[htbp]
\centering
\caption{Performance Comparison of SB, GA, SA and DBF \\ BF at $\theta$=$90^{\circ}$,$\phi$=$75^{\circ}$ with upper sidelobe suppression}
\vspace{-0.8\baselineskip}
\label{tab:performance}
\begin{tabular}{lcccc}

\textbf{Metric} & \textbf{SB} & \textbf{GA} & \textbf{SA} & \textbf{DBF}\\

Sidelobe Level (dB) & \textbf{-20.68} & -15.39 & -16.30 & -17.67\\
Directivity & 572.12 & \textbf{705.32} & 586.67 & 565.81\\
Time-to-Target (s) & \textbf{55} & 692,756 & 13,960 & --\\
Single-Run Time (s) & 55 & 1360 & 220 & \textbf{43}\\

\end{tabular}
\end{table}

\subsection{SB Optimizer Analysis}

\subsubsection{Parameter Settings} 
The careful selection of parameters significantly influences algorithm convergence. Fortunately, through our exploration, we have developed a set of broadly applicable parameter configurations.
We use a linearly increasing time-dependent pumping amplitude $p(t) = 0.01t$, as shown in Fig. \ref{evolution}(a). Other parameters are set as $K = 1$, $\Delta = 0.5$. 
When $p(t)$ gradually increases from zero to a sufficiently large value, each oscillator exhibits a bifurcation with two stable branches. Fig. \ref{evolution}(b) shows a bifurcation phenomenon optimized at a $5\times5$ array with a 1-bit encoding. Through brute force validation, the results after the bifurcation found the global optimum.

The step number of iterations significantly affects the solution. For instance, in the case of complex optimization problems (such as the BF with sidelobe suppression), more iterations are required to achieve a superior solution compared to simple optimization problems (such as the single beam).
Fig. \ref{evolution}(c) shows the optimization process of a $10\times24$ 3-bit phased array with sidelobe suppression and BF at ($\theta$ = $90^{\circ}$, $\phi$ = $60^{\circ}$). Here, we choose different $\xi_0$ that corresponds to the optimal solution of each problem. The energy over steps is averaged from multiple experiments and is depicted in the diagram. In the majority of cases, as the number of iterations increases, the energy decreases, indicating the result of a superior solution. At around 1,000 steps, the energy reaches convergence, and further increasing the number of steps would lead to significant time consumption without a substantial improvement in solution quality. 
\begin{figure}[htbp]
\centering 
\includegraphics[width=0.6\textwidth]{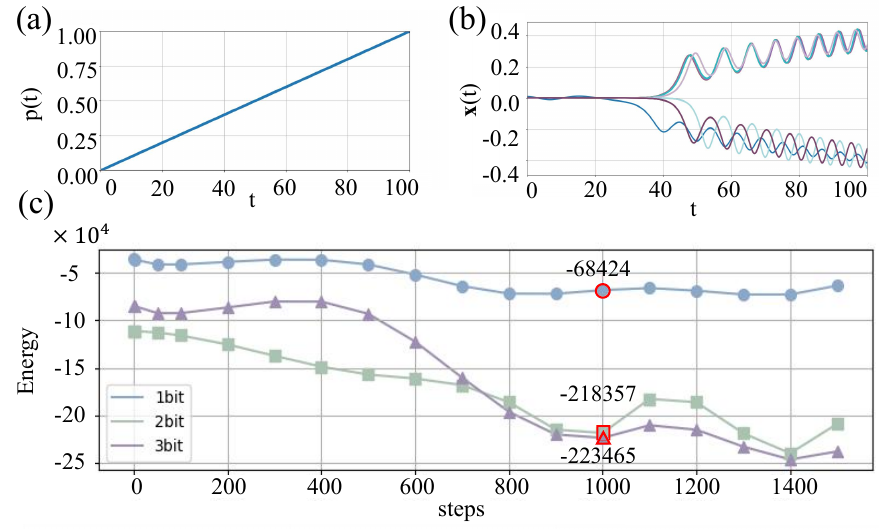}
\caption[The system.]{Time-dependent evolution process. (a): Pumping amplitude $p(t) = 0.01t$, meeting the requirement of adiabatic approximation condition. (b): bifurcation phenomenon of a $5\times5$ 1-bit phased array with time increasing. (c): Hamiltonian of a $10\times24$ 3-bit phased array with sidelobe suppression and BF at ($\theta$ = $90^{\circ}$, $\phi$ = $60^{\circ}$).  
}
\label{evolution}
\end{figure}

At times, the complexity of the problem may lead to an increase in energy when the number of iterations is increased. To strike a balance between solution quality and computational resources, we have opted for 1,000 steps as the default iteration count for the optimizer.
We could also observe that, in this case, the 2-bit result outperforms the 3-bit result within 800 steps. This is due to the larger search space for 3-bit solutions and smaller energy gap between near-ground states and the ground state (different energy landscape), which makes 3-bit optimization challenging. However, after surpassing 800 steps, with the use of the SB optimizer, the 3-bit result outperforms the 2-bit result, benefiting from the larger solution space provided by higher bits.

\subsubsection{Parameter Sensitivity} 
The parameter $\xi_0$ is the only one that cannot be determined. It significantly influences optimizer convergence and solution quality. Observing the sensitivity of the optimizer to $\xi_0$ is crucial.

We sampled $\xi_0$ from $10^{-8}$ to 1 with 2,000 exponentially spaced samples. The energy distribution for 1 to 3-bit results are illustrated in Fig. \ref{quality}. Fig. \ref{quality}(a) shows the energy distribution for the ($\theta$ = $50^{\circ}$, $\phi$ = $50^{\circ}$) single BF, Fig. \ref{quality}(b) displays the result of sidelobe suppression with BF at ($\theta$ = $90^{\circ}$, $\phi$ = $75^{\circ}$), and Fig. \ref{quality}(c) demonstrates the result of sidelobe suppression with BF at ($\theta$ = $90^{\circ}$, $\phi$ = $60^{\circ}$).

From the perspective of parameter sensitivity, we notice that different problems exhibit varying parameter sensitivities. For simple problems like Fig. \ref{quality}(a), nearly all $\xi_0$ values can find the optimal or near-optimal solution. However, for more complex problems like Fig. \ref{quality}(b) and Fig. \ref{quality}(c), the likelihood of finding optimal or near-optimal solutions through an extensive parameter scan decreases significantly. Comparing results with different rank of bit encoding, we observe that 1-bit encoding solutions are much worse than 2-bit and 3-bit encoding solutions. The 2-bit encoding solutions may offer comparable performance to 3-bit encoding solutions with greater stability, 
but the 3-bit solutions can break through the lowest energy of 2-bit solutions.
These findings corroborate the conclusions drawn in \textit{Parameter Settings} about the challenge caused by complex energy landscape.
\begin{figure*}[htbp]
\centering 
\includegraphics[width=0.95\textwidth]{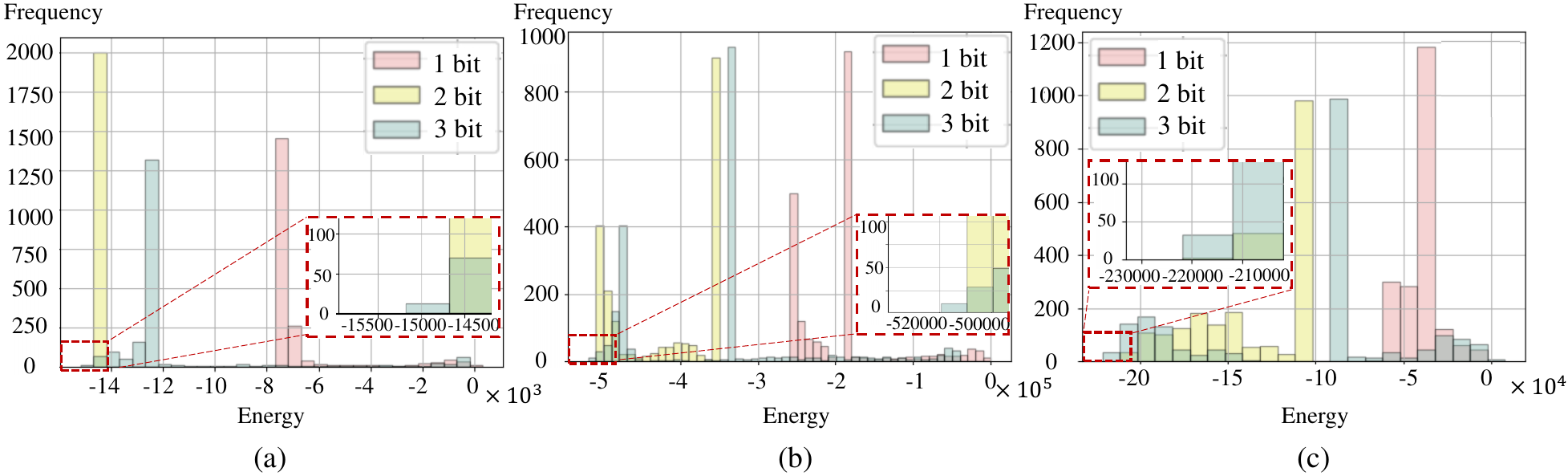}
\caption[The system.]{The energy distribution diagram obtained by exponentially sampling 2,000 groups of the parameter $\xi_0$ between $10^{-8}$ and 1. (a): single BF at ($\theta$ = $50^{\circ}$, $\phi$ = $50^{\circ}$). (b): sidelobe suppression with BF at ($\theta$ = $90^{\circ}$, $\phi$ = $75^{\circ}$). (c): sidelobe suppression with BF at ($\theta$ = $90^{\circ}$, $\phi$ = $60^{\circ}$).}
\vspace{-1\baselineskip}
\label{quality}
\end{figure*}

\subsubsection{Parallel Efficiency}

The efficiency of our approach is enhanced by two implemented levels of parallelism, supported by experimental validation.

\paragraph{Parallel Hamiltonian computation and reuse:}

For multi-objective optimization problems, the overall objective Hamiltonian is composed of a linear combination of several Hamiltonians. These Hamiltonians are independent of each other, allowing for parallel computation. Therefore, with parallel computing, the overall computation time does not increase linearly with the number of objectives. To illustrate this, in TABLE~\ref{tab:ham_time}, we present the computation times for serial and parallel computing across three proposed scenarios, each involving Hamiltonians with a varying number of objectives. A comparison between parallel and serial computation times demonstrates that parallel computing significantly reduces computation time for multi-objective optimization problems, with the improvement becoming more pronounced as the number of objectives increases.

\begin{table}[htbp]
\centering
\caption{Hamiltonian computation time with/without parallel}
\vspace{-0.8\baselineskip}
\label{tab:ham_time}
\begin{tabular}{lcc}
Scenario (3-bit quantization) & Sequential (s) & Parallel (s) \\
Single beam & 0.51 & \textbf{0.51} \\
Multi-beam (3 beams, 2 nulls) & 0.84 & \textbf{0.51} \\
Sidelobe suppression (1 beam, 171 nulls) & 16.85 & \textbf{0.53} \\
\end{tabular}
\end{table}

\paragraph{Parallel parameter tuning:}

\begin{figure}[htbp]
\centering 
\includegraphics[width=0.5\textwidth]{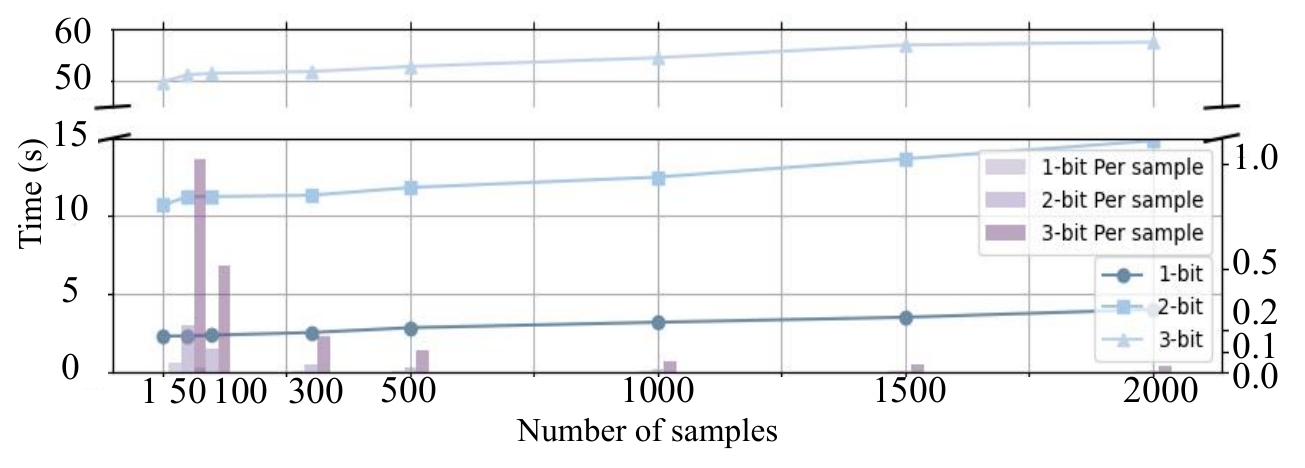}
\caption{Parameter tuning scalability: Total time (line) and average time/group (bars) vs. number of $\xi_0$ groups. Parallel parameter tuning capability enables simultaneous evaluation of up to 1,000 parameter configurations in a single run, thereby ensuring robust performance across diverse operational scenarios.}
\label{speed}
\end{figure}

The SB algorithm's performance is highly sensitive to parameter $\xi_0$, necessitating large-scale scanning. We implement parallel evaluation through matrix-based batch processing where each column represents a distinct $\xi_0$ value. 
Fig.~\ref{speed} demonstrates the strong scalability of our matrix-based parallel tuning strategy for critical parameter $\xi_0$. For example, processing 100 parameter groups increases total time by only 3\% (1-bit: 3 s $\rightarrow$ 3.1 s) while reducing the average time per group to 0.03s - achieving 97.3\% parallel efficiency. The 3-bit encoding maintains 50 to 60 s total time across scales, enabling 1,000-parameter scans in 54 s versus 13.9 h sequential computation.

\section{Conclusion}
This study presents significant advances in quantized phase-only massive MIMO BF optimization through the development of a quantum-inspired simulated bifurcation (SB) algorithm. The proposed method demonstrates clear advantages over conventional BF schemes, particularly in handling the complex combinatorial optimization challenges presented by quantized phase-only massive MIMO arrays.
The technical contributions of this research are multifaceted:

First, we developed a groundbreaking encoding technique that maps quantized phase values to spins while maintaining amplitude consistency, enabling precise 3-bit (and higher) phase optimization without the typical limitations of conventional methods. This encoding scheme, combined with our weighted Hamiltonian formulation, provides a physically intuitive framework for BF optimization that can be extended to other communication problems. Importantly, our implementation directly solves HUBO problems without introducing auxiliary spins, significantly reducing computational overhead compared to approaches that require problem quadratization.

Second, through extensive numerical experiments on a 240-element array, we validated the superior performance of our approach across multiple practical scenarios. 
The SB optimizer outperforms conventional approaches by overcoming the limitations of both semi-analytical methods in handling complex problems and stochastic optimization algorithms in prolonged optimization time and escaping local optima, demonstrating exceptional efficacy in challenging scenarios like single BF with sidelobe suppression.
A key engineering achievement is our parallel Hamiltonian calculation and parameter tuning strategy, which reduced Hamiltonian computation time for complex multi-objective optimization scenarios and enabled efficient scanning of thousands parameter groups within minutes. This computational efficiency, combined with our comprehensive parameter sensitivity analysis, makes the approach particularly valuable for real-world deployment where both solution quality and time-to-target are critical.

Third, our systematic investigation revealed crucial insights into the precision-performance tradeoff across different encoding schemes. As encoding precision increases from 1-bit to 3-bit, the Hamiltonian's energy landscape becomes progressively more complex. While this enhanced complexity enables discovery of higher-quality solutions, it also introduces optimization challenges through near-ground-state degeneracy effects. Our parallel computation strategy successfully mitigates these challenges, maintaining robust optimization performance even for the most demanding 3-bit encoding scenarios.

These advances are particularly impressive for next-generation wireless communication systems that demand both high precision and rapid adaptation, such as in drone swarm coordination and satellite-terrestrial integration. Future work will focus on adaptive control mechanisms to further accelerate convergence, hardware-specific optimizations for GPU implementations and further parallelization of SB algorithm via decoupled spin updates.

\newpage
\bibliographystyle{IEEEtran}
\bibliography{main.bib}

\begin{thebibliography}{10}
\providecommand{\url}[1]{#1}
\csname url@samestyle\endcsname
\providecommand{\newblock}{\relax}
\providecommand{\bibinfo}[2]{#2}
\providecommand{\BIBentrySTDinterwordspacing}{\spaceskip=0pt\relax}
\providecommand{\BIBentryALTinterwordstretchfactor}{4}
\providecommand{\BIBentryALTinterwordspacing}{\spaceskip=\fontdimen2\font plus
\BIBentryALTinterwordstretchfactor\fontdimen3\font minus \fontdimen4\font\relax}
\providecommand{\BIBforeignlanguage}[2]{{%
\expandafter\ifx\csname l@#1\endcsname\relax
\typeout{** WARNING: IEEEtran.bst: No hyphenation pattern has been}%
\typeout{** loaded for the language `#1'. Using the pattern for}%
\typeout{** the default language instead.}%
\else
\language=\csname l@#1\endcsname
\fi
#2}}
\providecommand{\BIBdecl}{\relax}
\BIBdecl

\bibitem{lu2014overview}
L.~Lu, G.~Y. Li, A.~L. Swindlehurst, A.~Ashikhmin, and R.~Zhang, ``An overview of massive mimo: Benefits and challenges,'' \emph{IEEE J. Sel. Top. Signal Process.}, vol.~8, no.~5, pp. 742--758, 2014.

\bibitem{paulraj2004overview}
A.~J. Paulraj, D.~A. Gore, R.~U. Nabar, and H.~Bolcskei, ``An overview of mimo communications-a key to gigabit wireless,'' \emph{Proc. IEEE}, vol.~92, no.~2, pp. 198--218, 2004.

\bibitem{hampton2013introduction}
J.~R. Hampton, \emph{Introduction to MIMO communications}.\hskip 1em plus 0.5em minus 0.4em\relax Cambridge university press, 2013.

\bibitem{tsoulos2018mimo}
G.~Tsoulos, \emph{MIMO system technology for wireless communications}.\hskip 1em plus 0.5em minus 0.4em\relax CRC press, 2018.

\bibitem{goldsmith2003capacity}
A.~Goldsmith, S.~A. Jafar, N.~Jindal, and S.~Vishwanath, ``Capacity limits of mimo channels,'' \emph{IEEE J. Sel. Areas Commun.}, vol.~21, no.~5, pp. 684--702, 2003.

\bibitem{haimovich2007mimo}
A.~M. Haimovich, R.~S. Blum, and L.~J. Cimini, ``Mimo radar with widely separated antennas,'' \emph{IEEE Signal Process Mag.}, vol.~25, no.~1, pp. 116--129, 2007.

\bibitem{ji2024electromagnetic}
R.~Ji, C.~Huang, X.~Chen, E.~Wei, L.~Dai, J.~He, Z.~Zhang, C.~Yuen, and M.~Debbah, ``Electromagnetic hybrid beamforming for holographic mimo communications,'' \emph{IEEE Transactions on Wireless Communications}, 2024.

\bibitem{kim2013multi}
C.~Kim, T.~Kim, and J.-Y. Seol, ``Multi-beam transmission diversity with hybrid beamforming for mimo-ofdm systems,'' in \emph{2013 IEEE Globecom Workshops (GC Wkshps)}.\hskip 1em plus 0.5em minus 0.4em\relax IEEE, 2013, pp. 61--65.

\bibitem{sun2014mimo}
S.~Sun, T.~S. Rappaport, R.~W. Heath, A.~Nix, and S.~Rangan, ``Mimo for millimeter-wave wireless communications: Beamforming, spatial multiplexing, or both?'' \emph{IEEE Commun. Mag.}, vol.~52, no.~12, pp. 110--121, 2014.

\bibitem{kulkarni2016comparison}
M.~N. Kulkarni, A.~Ghosh, and J.~G. Andrews, ``A comparison of mimo techniques in downlink millimeter wave cellular networks with hybrid beamforming,'' \emph{IEEE Trans. Commun.}, vol.~64, no.~5, pp. 1952--1967, 2016.

\bibitem{marzetta2015massive}
T.~L. Marzetta, ``Massive mimo: an introduction,'' \emph{Bell Labs Tech. J.}, vol.~20, pp. 11--22, 2015.

\bibitem{larsson2014massive}
E.~G. Larsson, O.~Edfors, F.~Tufvesson, and T.~L. Marzetta, ``Massive mimo for next generation wireless systems,'' \emph{IEEE Commun. Mag.}, vol.~52, no.~2, pp. 186--195, 2014.

\bibitem{marzetta2016fundamentals}
T.~L. Marzetta, E.~G. Larsson, and H.~Yang, \emph{Fundamentals of massive MIMO}.\hskip 1em plus 0.5em minus 0.4em\relax Cambridge University Press, 2016.

\bibitem{bjornson2016massive}
E.~Bj{\"o}rnson, E.~G. Larsson, and T.~L. Marzetta, ``Massive mimo: Ten myths and one critical question,'' \emph{IEEE Commun. Mag.}, vol.~54, no.~2, pp. 114--123, 2016.

\bibitem{molisch2017hybrid}
A.~F. Molisch, V.~V. Ratnam, S.~Han, Z.~Li, S.~L.~H. Nguyen, L.~Li, and K.~Haneda, ``Hybrid beamforming for massive mimo: A survey,'' \emph{IEEE Commun. Mag.}, vol.~55, no.~9, pp. 134--141, 2017.

\bibitem{ali2017beamforming}
E.~Ali, M.~Ismail, R.~Nordin, and N.~F. Abdulah, ``Beamforming techniques for massive mimo systems in 5g: overview, classification, and trends for future research,'' \emph{Front. Inf. Technol. Electron. Eng.}, vol.~18, pp. 753--772, 2017.

\bibitem{wu2018hybrid}
X.~Wu, D.~Liu, and F.~Yin, ``Hybrid beamforming for multi-user massive mimo systems,'' \emph{IEEE Trans. Commun.}, vol.~66, no.~9, pp. 3879--3891, 2018.

\bibitem{yang2018digital}
B.~Yang, Z.~Yu, J.~Lan, R.~Zhang, J.~Zhou, and W.~Hong, ``Digital beamforming-based massive mimo transceiver for 5g millimeter-wave communications,'' \emph{IEEE Trans. Microwave Theory Tech.}, vol.~66, no.~7, pp. 3403--3418, 2018.

\bibitem{maksymyuk2018deep}
T.~Maksymyuk, J.~Gazda, O.~Yaremko, and D.~Nevinskiy, ``Deep learning based massive mimo beamforming for 5g mobile network,'' in \emph{2018 IEEE 4th International Symposium on Wireless Systems within the International Conferences on Intelligent Data Acquisition and Advanced Computing Systems (IDAACS-SWS)}.\hskip 1em plus 0.5em minus 0.4em\relax IEEE, 2018, pp. 241--244.

\bibitem{black2017holographic}
E.~J. Black, ``Holographic beam forming and mimo,'' \emph{Pivotal Commware}, vol.~12, pp. 1--8, 2017.

\bibitem{litva1996digital}
J.~Litva and T.~K. Lo, \emph{Digital beamforming in wireless communications}.\hskip 1em plus 0.5em minus 0.4em\relax Artech House, Inc., 1996.

\bibitem{guo2017genetic}
H.~Guo, B.~Makki, and T.~Svensson, ``A genetic algorithm-based beamforming approach for delay-constrained networks,'' in \emph{2017 15th international symposium on modeling and optimization in mobile, ad hoc, and wireless networks (WiOpt)}.\hskip 1em plus 0.5em minus 0.4em\relax IEEE, 2017, pp. 1--7.

\bibitem{SA_2019_Ismayilov}
R.~Ismayilov, B.~Holfeld, R.~L.~G. Cavalcante, and M.~Kaneko, ``Power and beam optimization for uplink millimeter-wave hotspot communication systems,'' in \emph{2019 IEEE Wireless Communications and Networking Conference (WCNC)}, 2019, pp. 1--8.

\bibitem{ahmed2021machine}
I.~Ahmed, M.~K. Shahid, H.~Khammari, and M.~Masud, ``Machine learning based beam selection with low complexity hybrid beamforming design for 5g massive mimo systems,'' \emph{IEEE Trans. Green Commun. Networking}, vol.~5, no.~4, pp. 2160--2173, 2021.

\bibitem{liu2020machine}
X.~Liu, Y.~Liu, and Y.~Chen, ``Machine learning empowered trajectory and passive beamforming design in uav-ris wireless networks,'' \emph{IEEE J. Sel. Areas Commun.}, vol.~39, no.~7, pp. 2042--2055, 2020.

\bibitem{haupt1997phase}
R.~L. Haupt, ``Phase-only adaptive nulling with a genetic algorithm,'' \emph{IEEE Trans. Antennas Propag.}, vol.~45, no.~6, pp. 1009--1015, 1997.

\bibitem{deford1988phase}
J.~F. DeFord and O.~P. Gandhi, ``Phase-only synthesis of minimum peak sidelobe patterns for linear and planar arrays,'' \emph{IEEE Trans. Antennas Propag.}, vol.~36, no.~2, pp. 191--201, 1988.

\bibitem{morabito2012effective}
A.~F. Morabito, A.~Massa, P.~Rocca, and T.~Isernia, ``An effective approach to the synthesis of phase-only reconfigurable linear arrays,'' \emph{IEEE Trans. Antennas Propag.}, vol.~60, no.~8, pp. 3622--3631, 2012.

\bibitem{zeng2024performance}
Q.-G. Zeng, X.-P. Cui, B.~Liu, Y.~Wang, P.~Mosharev, and M.-H. Yung, ``Performance of quantum annealing inspired algorithms for combinatorial optimization problems,'' \emph{Communications Physics}, vol.~7, no.~1, p. 249, 2024.

\bibitem{krikidis2024optimizing}
I.~Krikidis, C.~Psomas, A.~K. Singh, and K.~Jamieson, ``Optimizing configuration selection in reconfigurable-antenna mimo systems: Physics-inspired heuristic solvers,'' \emph{IEEE Transactions on Communications}, 2024.

\bibitem{de2021materials}
N.~P. De~Leon, K.~M. Itoh, D.~Kim, K.~K. Mehta, T.~E. Northup, H.~Paik, B.~Palmer, N.~Samarth, S.~Sangtawesin, and D.~W. Steuerman, ``Materials challenges and opportunities for quantum computing hardware,'' \emph{Science}, vol. 372, no. 6539, p. eabb2823, 2021.

\bibitem{motta2022emerging}
M.~Motta and J.~E. Rice, ``Emerging quantum computing algorithms for quantum chemistry,'' \emph{Wiley Interdiscip. Rev.: Comput. Mol. Sci.}, vol.~12, no.~3, p. e1580, 2022.

\bibitem{hidary2019quantum}
J.~D. Hidary and J.~D. Hidary, \emph{Quantum computing: an applied approach}.\hskip 1em plus 0.5em minus 0.4em\relax Springer, 2019, vol.~1.

\bibitem{ross2021engineering}
C.~Ross, G.~Gradoni, Q.~J. Lim, and Z.~Peng, ``Engineering reflective metasurfaces with ising hamiltonian and quantum annealing,'' \emph{IEEE Trans. Antennas Propag.}, vol.~70, no.~4, pp. 2841--2854, 2021.

\bibitem{boothby2016fast}
T.~Boothby, A.~D. King, and A.~Roy, ``Fast clique minor generation in chimera qubit connectivity graphs,'' \emph{Quantum Information Processing}, vol.~15, pp. 495--508, 2016.

\bibitem{boothby2020next}
K.~Boothby, P.~Bunyk, J.~Raymond, and A.~Roy, ``Next-generation topology of d-wave quantum processors,'' \emph{arXiv preprint arXiv:2003.00133}, 2020.

\bibitem{willsch2022benchmarking}
D.~Willsch, M.~Willsch, C.~D. Gonzalez~Calaza, F.~Jin, H.~De~Raedt, M.~Svensson, and K.~Michielsen, ``Benchmarking advantage and d-wave 2000q quantum annealers with exact cover problems,'' \emph{Quantum Information Processing}, vol.~21, no.~4, p. 141, 2022.

\bibitem{goto2016bifurcation}
H.~Goto, ``Bifurcation-based adiabatic quantum computation with a nonlinear oscillator network,'' \emph{Sci. Rep.}, vol.~6, no.~1, p. 21686, 2016.

\bibitem{goto2019combinatorial}
H.~Goto, K.~Tatsumura, and A.~R. Dixon, ``Combinatorial optimization by simulating adiabatic bifurcations in nonlinear hamiltonian systems,'' \emph{Sci. Adv.}, vol.~5, no.~4, p. eaav2372, 2019.

\bibitem{goto2021high}
H.~Goto, K.~Endo, M.~Suzuki, Y.~Sakai, T.~Kanao, Y.~Hamakawa, R.~Hidaka, M.~Yamasaki, and K.~Tatsumura, ``High-performance combinatorial optimization based on classical mechanics,'' \emph{Sci. Adv.}, vol.~7, no.~6, p. eabe7953, 2021.

\bibitem{Biyingwang2023quantum}
B.-Y. Wang, H.~Ge, Y.~Jiang, S.~S.~A. Yuan, T.~Chu, Z.~Chen, S.~Pan, H.~Xu, G.~Zhang, X.~Cui, M.-H. Yung, F.~Liu, and W.~E.~I. Sha, ``Quantum-inspired optimization of beamforming with metasurfaces,'' in \emph{IEEE AP-S/URSI}, Portland, Oregon, USA, July 23–28, 2023.

\bibitem{obata2024ultra}
H.~Obata, T.~Nabetani, H.~Goto, and K.~Tatsumura, ``Ultra-high-speed optimization for 5g wireless resource allocation by simulated bifurcation machine,'' in \emph{2024 IEEE Wireless Communications and Networking Conference (WCNC)}.\hskip 1em plus 0.5em minus 0.4em\relax IEEE, 2024, pp. 01--06.

\bibitem{zaman2021pyqubo}
M.~Zaman, K.~Tanahashi, and S.~Tanaka, ``Pyqubo: Python library for mapping combinatorial optimization problems to qubo form,'' \emph{IEEE Trans. Comput.}, vol.~71, no.~4, pp. 838--850, 2021.

\bibitem{glover2018tutorial}
F.~Glover, G.~Kochenberger, and Y.~Du, ``A tutorial on formulating and using qubo models,'' \emph{arXiv preprint arXiv:1811.11538}, 2018.

\bibitem{pastorello2019quantum}
D.~Pastorello and E.~Blanzieri, ``Quantum annealing learning search for solving qubo problems,'' \emph{Quantum Inf. Process.}, vol.~18, no.~10, p. 303, 2019.

\bibitem{schultz1964two}
T.~D. Schultz, D.~C. Mattis, and E.~H. Lieb, ``Two-dimensional ising model as a soluble problem of many fermions,'' \emph{Rev. Mod. Phys.}, vol.~36, no.~3, p. 856, 1964.

\bibitem{dupuis197916}
P.~Dupuis, M.~Joindot, A.~Leclert, and D.~Soufflet, ``16 qam modulation for high capacity digital radio system,'' \emph{IEEE Trans. Commun.}, vol.~27, no.~12, pp. 1771--1782, 1979.

\bibitem{chong2003new}
C.~V. Chong, R.~Venkataramani, and V.~Tarokh, ``A new construction of 16-qam golay complementary sequences,'' \emph{IEEE Trans. Inf. Theory}, vol.~49, no.~11, pp. 2953--2959, 2003.

\bibitem{kim2019leveraging}
M.~Kim, D.~Venturelli, and K.~Jamieson, ``Leveraging quantum annealing for large mimo processing in centralized radio access networks,'' in \emph{Proceedings of the ACM special interest group on data communication}, 2019, pp. 241--255.

\bibitem{balanis2016antenna}
C.~A. Balanis, \emph{Antenna theory: analysis and design}.\hskip 1em plus 0.5em minus 0.4em\relax John wiley \& sons, 2016.

\bibitem{gottlieb1998total}
S.~Gottlieb and C.-W. Shu, ``Total variation diminishing runge-kutta schemes,'' \emph{Math. Comput.}, vol.~67, no. 221, pp. 73--85, 1998.

\bibitem{boros2002pseudo}
E.~Boros and P.~L. Hammer, ``Pseudo-boolean optimization,'' \emph{Discrete Appl. Math.}, vol. 123, no. 1-3, pp. 155--225, 2002.

\bibitem{jiang2018quantum}
S.~Jiang, K.~A. Britt, A.~J. McCaskey, T.~S. Humble, and S.~Kais, ``Quantum annealing for prime factorization,'' \emph{Sci. Rep.}, vol.~8, no.~1, p. 17667, 2018.

\bibitem{mato2022quantum}
K.~Mato, R.~Mengoni, D.~Ottaviani, and G.~Palermo, ``Quantum molecular unfolding,'' \emph{Quantum Sci. Technol.}, vol.~7, no.~3, p. 035020, 2022.

\bibitem{kanao2022simulated}
T.~Kanao and H.~Goto, ``Simulated bifurcation for higher-order cost functions,'' \emph{Appl. Phys. Express}, vol.~16, no.~1, p. 014501, 2022.

\bibitem{Mirjalili2019}
\BIBentryALTinterwordspacing
S.~Mirjalili, \emph{Genetic Algorithm}.\hskip 1em plus 0.5em minus 0.4em\relax Cham: Springer International Publishing, 2019, pp. 43--55. [Online]. Available: \url{https://doi.org/10.1007/978-3-319-93025-1_4}
\BIBentrySTDinterwordspacing

\bibitem{Vecchi_SA}
\BIBentryALTinterwordspacing
S.~Kirkpatrick, C.~D. Gelatt, and M.~P. Vecchi, ``Optimization by simulated annealing,'' \emph{Science}, vol. 220, no. 4598, pp. 671--680, 1983. [Online]. Available: \url{https://www.science.org/doi/abs/10.1126/science.220.4598.671}
\BIBentrySTDinterwordspacing

\end{thebibliography}

\end{document}